\documentclass[preprint,aps,superscriptaddress,showkeys,11pt]{revtex4}
\pdfoutput=1
\usepackage{graphicx}
\usepackage{amsmath}
\usepackage{amsfonts}
\usepackage{amssymb}
\usepackage{xcolor, soul}
\usepackage{epstopdf}
\usepackage{physics}
\usepackage{float}
\usepackage{subfigure}
\usepackage{hyperref}
\hypersetup{
     colorlinks   = true,
     citecolor    = red,
     linkcolor    = blue,
     urlcolor     = blue,
}
\def\beq{\begin{equation}}
\def\eeq{\end{equation}}
\def\bea{\begin{eqnarray}}
\def\eea{\end{eqnarray}}
\def\be{\begin{equation}}
\def\ee{\end{equation}}

\def\nno{\nonumber}
\def\bse{\begin{subequations}}
\def\ese{\end{subequations}}

\def\tR{\tilde{R}}
\def\tV{\tilde{V}}
\def\tg{\tilde{g}}

\newcommand{\phiast}{\phi_{\ast}}

\graphicspath{{./figs/}}

\begin{document}

\title{Minimal plateau inflationary cosmologies and constraints from reheating}

\author{Debaprasad Maity}
 \email{debu@iitg.ac.in}
\author{Pankaj Saha}%
 \email{pankaj.saha@iitg.ac.in}
\affiliation{%
Department of Physics, Indian Institute of Technology Guwahati.\\
 Guwahati, Assam, India 
}%

\date{\today}

\begin{abstract}
With the growing consensus on simple power law inflation models not being favored by the PLANCK observations, dynamics for the non-standard inflation gain significant interest in the recent past. In this paper, we analyze in detail a class of supergravity inspired phenomenological inflationary models with non-polynomial potential based on\cite{mhiggs}, and compare the model predictions with the currently most favored Starobinsky and its generalized $\alpha$-attractor models in the $(n_s,r)$ plane constrained by PLANCK. Importantly for a wide range of parameter space, our model provides successful inflation in the sub-Planckian regime. We also have performed model independent analysis of reheating in terms of the effective equation of state parameter. In particular, we consider two stages of reheating dynamics with generalized inflaton equation of state in the initial and relativistic equation of state in the later phase. Finally, we show how our generalized reheating analysis constrains the inflation models under consideration.   
\end{abstract}

\maketitle

\tableofcontents

\section{\label{intro}Introduction}
The inflation\cite{guth,Starobinsky:1980te,Sato:1980yn,linde1,steinhardt} is a model independent mechanism proposed to solve some of the outstanding problems in standard Big-Bang cosmology. It is an early exponential expansion phase of our universe, which sets the required initial conditions for the standard Big-Bang evolution. Over the years a large number of models have been introduced to realize this mechanism\cite{Martin:2013tda}, and to explain the cosmological observations\cite{PLANCK}. The latest PLANCK data turned out to favor inflationary models with plateau potential. Such kind of plateau models include the well known Starobinsky model\cite{Starobinsky:1980te}, the Chaotic inflation in supergravity \cite{Goncharov:1983mw}, the Higgs inflation\cite{Salopek:1988qh,Bezrukov:2007ep}. A particularly interesting class of models that also unifies and generalizes a broad class of plateau models are the $\alpha$ attractor models\cite{Kallosh:2013hoa,Ferrara:2013rsa,Kallosh:2013daa,Kallosh:2013yoa,Galante:2014ifa}. In this paper, we will introduce a new class of inflationary model generalizing the work proposed in \cite{mhiggs}. We also discussed two possible ways of generating such plateau potentials. Our first approach is to realize the potential in the supergravity framework. We have shown that choosing particular forms of supergravity potentials we can generate the scalar field plateau potential. However, important ingredient in our construction is the presence of anomalous $U(1)$ symmetry. This supergravity inspired scalar potentials will reduce to our desired power-law plateau potentials in a specific limit. We have also provided another realization of our potential starting from the simple power-law potentials $V(\phi) \propto \phi^n$ with general scalar-tensor theory given in appendix-\ref{appendix:appA}. At this point let us motivate the reader pointing out important points of our study. It is well known that the simple power-law canonical potentials of the form $V(\phi) \propto |\phi|^n$ are not cosmologically viable because of their prediction of large tensor-to-scalar ratio. In addition, because of super-Planckian initial value of the scalar field, the effective field theory description  may not be valid. One of our goals in this paper is to circumvent the above mentioned problems in the framework of canonical scalar field theory. In order to achieve this, we generalize the power-law potentials to a non-polynomial form such that it can give successful inflation with sub-Plankian initial condition. After the inflation, the production of radiation and other matter fields occur during reheating phase which also sets the initial condition for the standard big-band. Therefore, for completeness we also study the reheating phase considering the simplest scenario where inflaton is decaying into radiation though discrete change of equation of state \cite{kamionkowski}.  It is well known that for inflation with the potential $V(\phi) \propto |\phi|^n$ the effective equation of state can be defined as $w_{\rm eff}=(n-2)/(n+2)$\cite{Turner:1983he,Mukhanov:2005sc,Martin:2010kz}. As the usual power-law potentials for $n\geq2$ turned out to be disfavored from CMB data, a detailed analysis of this phase for arbitrary power-law inflaton potential, to the best of our knowledge, is still missing. Therefore, in this paper for the first time we generalize the existing reheating constraint analysis by considering the above general inflaton equation of state, and qualitatively include the fact that $w_{eff}$ has to go to that of radiation because of inflaton decay at the end of reheating. This is what we call two stage reheating. We believe that our present two stage reheating approach is more realistic compared to that of usual reheating constraint analysis proposed in \cite{kamionkowski}. However detail studies of the perturbative and non-perturbative issues of reheating phase for general equation of state will be reported elsewhere. To complete our discussion, in the appendix-\ref{appendix:appB}, we briefly discuss about an important theoretical issue related to unitarity. Since our model has an additional scale $\phi_*$ which controls the dynamics and sets the scale of inflation, it is very important to maintain the unitarity scale say $\Lambda$ to be greater than $\phi_*$ during the inflation period. 

We structured our paper as follows: In section-\ref{model}, we generalize the model introduced in \cite{mhiggs}, and study in detail the cosmological dynamics of inflaton starting from inflation to reheating. We compute the important cosmological parameters such as scalar spectral index $(n_s)$, the tensor to scalar ratio $(r)$, and the spectral running $(\dd{n_s^k})$ and compare them with the observations. From those cosmological observations, we constrain the parameters of our models. In section-\ref{sugra}, we constructed the supergravity realization of our model and also compare the PLANCK result with the model under consideration. After the end of inflation, in general, the inflaton starts to have coherent oscillation around the minimum of the potential, during which the universe said to undergo reheating phase. We have computed the effective equation of state of the oscillating inflaton to study the reheating phase. In section-\ref{reheatingprediction}, we have done the model independent analysis for possible ranges of reheating temperatures $(T_{re})$, and e-folding numbers $(N_{re})$ during reheating considering the background expansion and the evolution of entropy density. These considerations put further constraints on the parameter space of our model. We concluded in section-\ref{conclusion}. 

We will consider $\hbar = c = 1$ unless otherwise stated. We have denoted ${\rm M_p}( = 1/\sqrt{8\pi G}) = 2.43\times 10^{18}{\rm GeV}$ as the reduced Planck constant. We will take the usual Friedmann-Le\^{i}matre-Roberson-Walker (FLRW) metric as our background metric $\dd{s}^2= \dd{t}^2 -a^2(t)(\dd{x}^2+\dd y^2+\dd{z}^2) $ for deriving our equations. Where $a(t)$ is the scale factor and $t$ represents the cosmic time.

\section{\label{model}The Model}
In this section will describe a class of phenomenological power-law plateau potentials which is flat at large field values. The form of the potentials are given as,
\bea
V_{\rm min}(\phi) = \lambda \frac{m^{4-n} \phi^n}{1+\left(\frac{\phi}{\phi_*}\right)^n} . 
\eea
In the above form of the potentials, the parameters $n$, and $\lambda$ or $m$ has the same role as in chaotic power-law inflationary models. The index $n$ is assumed to be even integer. The parameter $\lambda$ assumes non-trivial value only for $n=4$, which has been studied as minimal Higgs inflation in \cite{mhiggs}. For other values of $n$, we will set $\lambda =1$. In our model we have introduced a mass scale $\phiast$ which controls the shape of the potentials. For large field value the potentials becomes flat which sets  the scale of inflation as $\Lambda = \lambda m^{4-n} \phi_*^n$. One can further generalize this model by considering the potential to be dependent only upon the modulus of the inflaton field. In that case, $n$ can take all positive integer value. For the sake of simplicity we will stick to only even values of $n$. Another simple generalization of our model can be done by defining $V(\phi)^q$ as a new potential. Where, $q$ will be an another positive integer. As emphasized before, in section-\ref{sugra}, we will describe possible realization of these type of plateau potentials in the supergravity framework specifically in the sub-Planckian limit of $\phi_*$. An alternative construction of this class of potential from general scalar-tensor theory has also been presented in the appendix-\ref{appendix:appA}.


\subsection{\label{background}Background Equations}
In this section we will study the background dynamics using the above form of the potentials. We will start with the following action,
\bea
S ~=~ \int \dd^4x \sqrt{-g} \left[\frac {\rm M_p^2}{2}  R - \frac{1}{2}g^{\mu\nu}\partial_{\mu}\phi \partial_{\nu}\phi- V_{\rm min}(\phi) \right]
\label{action}
\eea
Assuming the usual FLRW background ansatz for the space-time, 
the system of equations governing the dynamics of inflaton and scale factor are
\begin{align}
3{\rm M_{p}^2} H^2 &= \frac{1}{2}\dot{\phi}^2 + V_{\rm min}(\phi)\\
2{\rm M_{p}^2} \dot{H} &= -\dot{\phi}^2\\
\ddot{\phi} + 3 H \dot{\phi} + V'_{\rm min}(\phi) &= 0 ,
\label{friedman}
\end{align}
where, the usual definition of Hubble constant is $H = {\dot{a}}/a$.
As the potential is asymptotically flat for large field value compared to $\phi_*$, the condition for sufficient inflation is automatically satisfied. The flatness conditions for the potential during inflation are quantified in terms of the \textit{slow-roll} parameters, defied as 
 \bea
   \epsilon \equiv \frac{\rm M_p^2}{2} \left( \frac{V'_{\rm min}}{V_{\rm min}}\right)^2 &=&
   \nno
      \frac{n^2 {\rm M_p^2} \phi_* ^{2 n}}{2 \phi ^2 \left(\phi_* ^n+\phi ^n\right)^2} \\        
               \eta \equiv {\rm M_p^2} \left( \frac{V''_{\rm min}}{V_{\rm min}}\right) &=&
          \frac{n {\rm M_p^2} \phi_* ^n \left((n-1) \phi_* ^n-(n+1) \phi ^n\right)}{\phi ^2 \left(\phi_* ^n+\phi ^n\right)^2} .
  \label{eq-slow-roll}
\eea
During inflation $\epsilon \ll 1$ and $|\eta| \ll 1$. Therefore, the end of inflation 
is usually set by the condition $\epsilon=1$. Let us also define the third order slow-roll parameter related to the third derivative of the potential as,
\be
\xi \equiv {\rm M_p^4} \left(\frac{V'_{\rm min} V'''_{\rm min}}{V^2} \right) =
\frac{n^2 {\rm M_p^4} \phi_* ^{2 n} \left(\left(n^2-3 n+2\right) \phi_* ^{2 n}-4 \left(n^2-1\right) \phi_* ^n 
\phi ^n+\left(n^2+3 n+2\right) \phi ^{2 n}\right)}{\phi ^4 \left(\phi_* ^n+\phi ^n\right)^4} .
\ee
In addition to provide the successful inflation, all the aforementioned slow-roll 
parameters play very important role in controlling the dynamics of cosmological perturbations during inflation. An important cosmological parameter, which quantifies the amount of inflation is called e-folding number $(N)$. The e-folding number is expressed as
\bea
N = \ln\left(\frac{a_{\rm end}}{a_{\rm in}}\right) = \int\limits_{a_{\rm in}}^{a_{\rm end}} \dd \ln a = \int\limits_{t_{\rm in}}^{t_{\rm end}} H \dd{t}   \simeq 
 \int\limits_{\phi_{\rm in}}^{\phi_{\rm end}} \frac{1}{\sqrt{2 \epsilon}} \frac{|\dd \phi|}{\rm M_p} .
\label{efold}
\eea
 As we have mentioned, the inflation ends when $\epsilon=1$, therefore, one can use  Eq.(\ref{efold}) to find the value of the inflaton field when a particular mode exits the horizon during inflation. 
By solving the aforementioned condition, we can express the e-folding number $N$ into the following form,  
    \begin{align}
       N=\frac{\phi_*^2}{n {\rm M_p}^2}\left[\frac{1}{(n+2)}(\tilde{\phi}^{(n+2)}-\tilde{\phi}_{end}^{(n+2)}) + \frac{1}{2} (\tilde{\phi}^2-\tilde{\phi}_{end}^2)\right] &  \simeq \frac{\phi_*^2}{n {\rm M_p}^2} \frac{1}{(n+2)} \tilde{\phi}^{(n+2)}.
       \label{efold3}
     \end{align}
Where we have defined, $\tilde{\phi} = {\phi}/{\phi_*}$. In the above expressions for $N$, we have ignored the contribution coming from $\phi_{\rm end}$, and its squared terms. 
We have numerically checked the validity of those expressions for a wide range
of value of $\phi_* \leq {\cal O}({\rm M_p})$. From cosmological observations one needs $N \simeq 50-60$, so that the scales of our interest in CMB were inside the causal horizon during inflation. By using the above mentioned boundary conditions for the inflaton we have solved for the homogeneous part of inflaton $\phi(t)$ field and the scale factor $a(t)$. Next we describe the relevant cosmological parameters associated various correlation functions of the fluctuation.    

\subsection{\label{pert}  Computation of $(n_s,r,\dd{n_s^k})$}
 As we have described in the introduction, the very idea of inflation was introduced to solve 
the outstanding problems of standard Big-Bang cosmology. Soon it was realized that inflation also provides seed for the large-scale structure of our universe through quantum fluctuation. All the cosmologically relevant inflationary observables are identified with various correlation functions of those primordial fluctuations calculated in the framework of quantum field theory. 

\begin{figure}
 \includegraphics[scale=1.0]{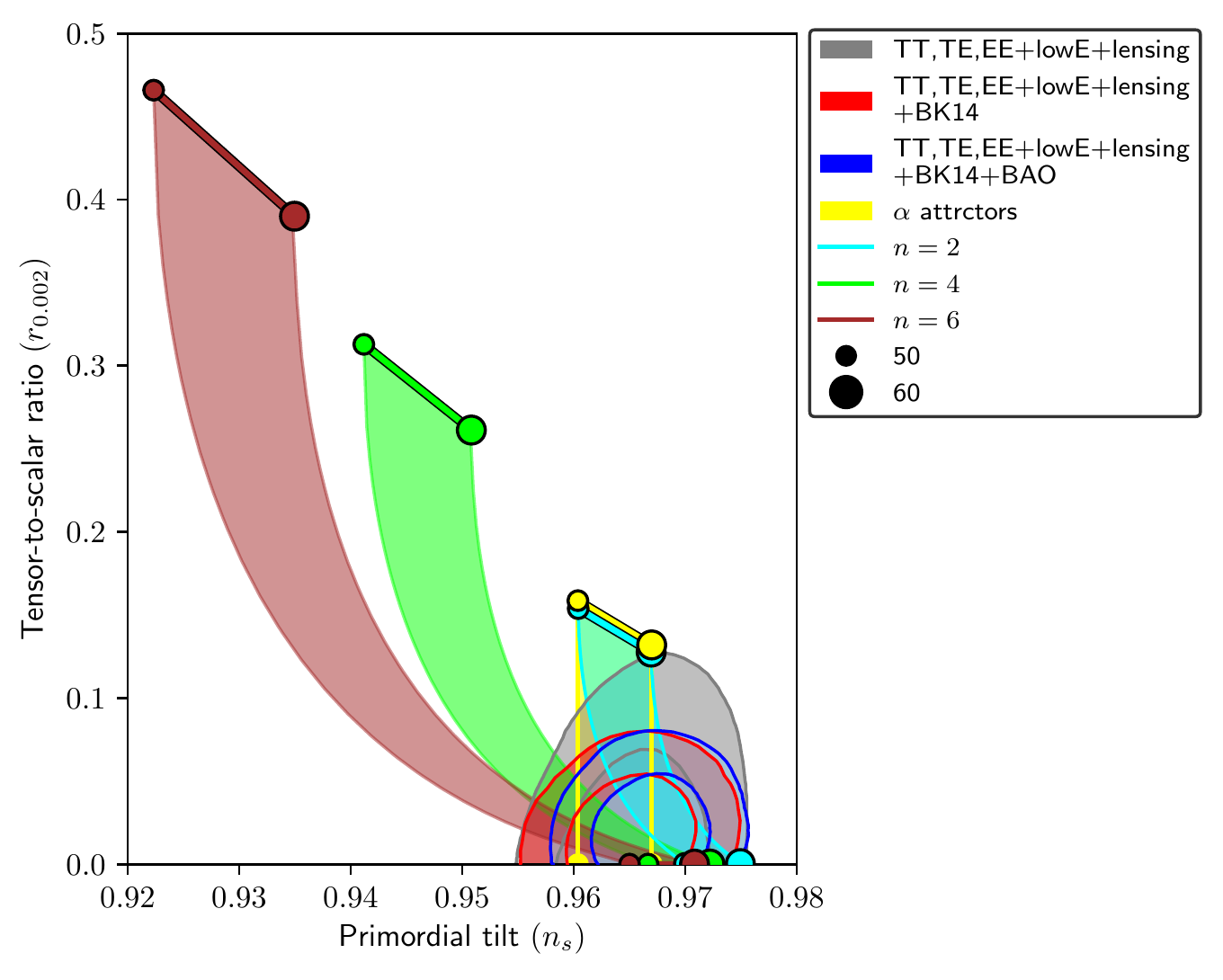}
\caption{\scriptsize The $n_s$-$r$ plot of the model on the marginalized joint 68\% and 95\% CL regions at $k = 0.002Mpc^{-1}$ from Planck alone and in combination with BK14 or BK14 plus BAO data\cite{Akrami:2018odb}.}
\label{planckplot}
\end{figure}

We have curvature and tensor perturbation. The two and higher point correlation functions of those fluctuations are parametrized by power spectrum(see, \cite{cpt1,cpt2,Baumann:2009ds,Baumann:2018muz} for a comprehensive review of Cosmological Perturbation Theory). The scalar curvature power spectrum is given by
 \be
 \Delta_{\mathcal{R}}^2 = \frac{1}{8 \pi^2} \frac{1}{\epsilon} \frac{H^2}{\rm M_p^2} \bigg|_{k=aH} = \frac{1}{12 \pi^2} \frac{V_{\rm min}^3}{{\rm M_p^6} (V'_{\rm min})^2} .
 \ee
Once we know the power spectrum, the cosmological quantity of our interests are the spectral tilt and its running. During inflation a particular inflaton field value corresponds to a particular momentum mode exiting the horizon. Hence by using the following relation to the leading order
in slow-roll parameters, $\dv{}{\ln k} = \frac{\dot{\phi}}{H}\frac{d}{\dd\phi}$, one 
obtains the following inflationary observables
\bea
n_s -1 \equiv \dv{\ln \Delta_{\cal R}^2}{\ln k} = -6\epsilon + 2 \eta
\eea
\bea
\dd n_s^k \equiv \dv{n_s}{\ln k} = -2 \xi + 16 \epsilon \eta - 24 \epsilon^2 .
\eea
Similarly we can compute the tensor power spectrum $\Delta_{t}$ for the gauge invariant tenor perturbation $h_{ij}$ that generates the primordial gravitational waves. Finally normalizing the gravitational wave amplitude with the scalar one, we get the scalar-to-tensor ratio 
\bea
r = \frac{\Delta_{t}^2}{\Delta_{\cal R}^2} = 16\epsilon .
\eea
Once we have all the expression for cosmological quantities in terms of slow roll parameters, by using Eqs.(\ref{eq-slow-roll}) and (\ref{efold3}), and considering $\phi_* \leq {\cal{O}}(1)$ in unit of ${\rm M_p}$, we express $(n_s,r,\dd{n_s^k})$ in terms of $n$,$N$, and $\phi_*$, as 
 \bea \label{nsrvsN}
 1 - n_s &=& 
 \frac{2(n+1)}{(n+2)} \frac{1}{N}
  ~~;~~dn_s^k = 
-\frac{(2+3n+n^2)}{(n+2)^2} \frac{1}{N^2}
\\ \nno
  r &=& 
  8n^2 \left(\frac{\phi_*}{\rm M_p}\right)^{\frac{2n}{(n+2)}} \frac{1}{[n(n+2)]^{\frac{2(n+1)}{(n+2)}} N^{\frac{2(n+1)}{(n+2)}}} .
\eea
The numerical results for inflationary predictions of this model is shown in Fig.(\ref{planckplot}) with respect to the latest Planck data\cite{PLANCK}.
It is evident that the above analytic expressions for the inflationary observables derived in Eq.(\ref{nsrvsN}) are valid for subplanckian values of the scale $\phi_{\ast}$. For large values of this scale, the potentials reduce to the simple power-law form $V(\phi)\propto \phi^n$ and the observables are given by well-known results
\begin{align}
 1-n_s = \frac{(2+n)}{2N},\quad r=\frac{4n}{N},\quad \dd n^k_s = -\frac{2}{N^2},~~
 \label{eq:largeNchaotic}.
\end{align}
At this point let us point out an important difference between our models and the $\alpha$-attractor models. For large $N$, with $\alpha \ll N$, the spectral data for the $\alpha$-attarctor models are\cite{Galante:2014ifa}
\begin{equation}
 1-n_s = \frac{2}{N},\quad r = \frac{12\alpha}{N^2}.
 \label{eq:largeNaplpha}
\end{equation}
This is in sharp contrast with that of our models given in Eq.(\ref{nsrvsN}). Therefore, as emphasized before, our model will not fall in the class of $\alpha$-attractor models. An important feature of the $\alpha$-attractor models is that the cosmological parameters do not depend upon the details of the potentials specifically on the power of the potential. We will do the detailed comparison on this in section \ref{sugra}.

So far all we have discussed is directly related to the cosmological observation made by PLANCK. Another important quantity of theoretical interest we would like to compute is Lyth bound \cite{lythbound} $\Delta \phi$. 
This quantity measures the difference of field values which is traversed by the inflaton
field during inflation. This is so calculated that for a particular model $\Delta \phi$ is the maximum possible
value for a particular e-folding number. Inflation is a semi-classical phenomena. It is believed that natural cut off scale for any theory
minimally or non-minimally coupled with gravity is Planck scale ${\rm M_p}$. Therefore, amount of inflaton field value 
can naturally be a good measure to tell us the effective validity of a model under study in the effective field theory
language. Hence the calculated expression for the field excursion in terms of $N$ and $\phi_*$ are:
\bea
\Delta \phi \gtrsim {\rm M_p} N \sqrt{\frac{r}{8}} = 
{\rm M_p} \left(  \frac{n}{n+2} \right) \frac{1}{\left[n(n+2)\right]^{\left(\frac{n+1}{n+2}\right)}} N^{\frac{1}{(n+2)}}
\eea
\begin{table}[t]
	\begin{tabular}{|p{0.7cm}|p{0.6cm}|p{1.0cm}|p{1.5cm}| p{1.5cm} |p{1.0cm}|}
		\hline
		$\frac{\phi_*}{\rm M_p}$ & $n$ & $n_s$ &  $r$ & $dn_s^k$  &  $\Delta \phi $ 
		\\
		\hline
		\hline
		0.01    &  2  & 0.969 & $4\times10^{-5}$ & -0.00066 & 0.39 \\
		\cline{2-6}
		&  4  & 0.966 &  $2\times10^{-6}$ & -0.00066 & 0.12 \\
		\cline{2-6}
		&  6  & 0.965 & $3\times10^{-7}$ & -0.00069 & 0.06  \\
		\hline
		\hline
		1.00    &  2  & 0.969 & $4\times10^{-3}$ & -0.0006 & 3.53  \\
		\cline{2-6}
		&  4  & 0.966 & $9.6\times10^{-4}$ & -0.0007 & 2.13  \\  
		\cline{2-6}
		&  6  & 0.964 & $3.5\times10^{-4}$ & -0.0007 & 1.47\\
		\hline
		\hline
	\end{tabular}
	\caption{\scriptsize The spectral quantities for different values of $n$ for 50 efolding. The two values of $\phi_*$ are chosen to illustrate that we can have both small field and large field inflation 
		depending on the value of $\phi_*$.}
	\label{tab1}
\end{table}
All the quantities we have discussed so far is independent of $m$ or $\lambda$. (At this point let us again remind the reader that for $n \neq 4$, $\lambda$ is a dimensionless quartic coupling parameter. While for $n\neq 4$, $m$ is dimensionful parameter, and we set $\lambda =1$). However, comparing the inflationary power spectrum with the PLANCK normalization we will determine the value of $m$ or $\lambda$ and then calculate all the other quantities of our interest. The expression for the power spectrum of the curvature perturbation is 
\bea
\Delta_{\cal R}^2 =
\frac{\lambda}{12 \pi^2 n^2} \left(\frac{m}{\rm M_p}\right)^{4-n} \left(\frac{\phi_*}{\rm M_p}\right)^{\frac{n^2}{n+2}} \left[n(n+2) N\right]^{\frac{2(n+1)}{(n+2)}} = 2.4 \times 10^{-9}.
\label{powspectrum}
\eea 
As mentioned we considered the PLANCK normalization: $\Delta_{\cal R}^2$ at the pivot scale $k/a_0 = 0.05 Mpc^{-1}$, and corresponding estimated scalar spectral index is $n_s = 0.9682 \pm 0.0062$. 
\begin{table}[t]
	\begin{tabular}{|p{1.2cm}|p{0.6cm}|p{1.6cm}|p{1.6cm}| p{1.6cm} |p{1.6cm}|}
		\hline
		$\phi_*/{\rm M_p}$ & $n$  & $m/{\rm M_p}$ &  $\lambda$ & $H_*/{\rm M_p}$  &  $V_*^{1/4}/{\rm M_p}$\\
		\hline
		\hline
		$0.01$   &  2  & $1.2\times10^{-4}$ & $1$ & $6.8\times10^{-7}$ & $1.1\times10^{-3}$  \\
		\cline{2-6}
		&  4  & - & $7.3\times10^{-6}$ & $1.6\times10^{-7}$ & $5.2\times10^{-4}$  \\
		\cline{2-6}
		&  6  & $8.9$ & 1 & $6.5\times10^{-8}$ & $3.3\times10^{-4}$ \\
		\cline{2-6}
		&  8  & $4.0\times10^{-1}$ & 1 & $3.6\times10^{-8}$ & $2.5\times10^{-4}$ \\
		\hline
		\hline
		$1$    &  2  & $1.3\times10^{-5}$ & $1$ & $4.6\times10^{-6}$ & $2.8\times10^{-3}$ \\
		\cline{2-6}
		&  4  & - &  $3.5\times10^{-11}$ & $2.7\times10^{-6}$ & $2.1\times10^{-3}$  \\
		\cline{2-6}
		&  6  & $2.8\times10^{5}$ & 1 & $1.7\times10^{-6}$ & $1.7\times10^{-3}$  \\
		\cline{2-6}
		&  8  & $6.4\times10^{2}$ & $1$ & $1.3\times10^{-6}$ & $1.5\times10^{-3}$   \\
		\hline
		\hline
		10    &  2  & $5.7\times10^{-6}$ & $1$ & $2.5\times10^{-5}$ & $6.5\times10^{-3}$ \\
		\cline{2-6}
		&  4  & - &  $1.5\times10^{-13}$ & $2.0\times10^{-5}$ & $6.0\times10^{-3}$  \\
		\cline{2-6}
		&  6  & $3.8\times10^{7}$ & 1 & $1.4\times10^{-5}$ & $5.0\times10^{-3}$  \\
		\cline{2-6}
		&  8  & $2.3\times10^{4}$ & $1$ & $1.0\times10^{-5}$ & $4.2\times10^{-3}$   \\
		\hline
		\hline
	\end{tabular}
	\caption{\scriptsize Inflationary energy scales and the parameter $m$ and $\lambda$ for two different values of $\phi_*$}
	\label{scales}
\end{table}

After having all our necessary expressions for the cosmological quantities, we have plotted our main results in $(n_s,r)$ plane and compared it with the experimental values $n_s=0.968 \pm 0.006$ and the upper limit on $r<0.11$ in the Fig.(\ref{planckplot}). In the 
table-(\ref{tab1}) we have given some sample values of all the cosmologically relevant quantities for different values of theoretical parameters.  As emphasized before, specifically for $\phi_*<\mbox{M}_p$, our model predictions match well with the PLANCK data in the sub-Planckian regime. Importantly we can envision infinite numbers of model potentials for different values of $(n,\phiast)$ which give rise to low scale inflation. Most interesting case would probably be $n=4$. In the recent paper \cite{mhiggs}, it has been identified as a minimal Higgs inflation. Even though this identification is not straight forward, however, for small field value Taylor expanding the potential one can identify $\lambda$ as Higgs quartic coupling which can be set to its electroweak value. Renormalization group analysis needs to be performed from inflation scale to the electroweak scale to make this identification precise. In addition  an important question needs to be addressed with regard to the unitarity of the models at the inflationary. For completeness, in the appendix we provide a short discussion on this issue, and details will be studied elsewhere. The values of $m$ and $\lambda$ for different models have been listed in Table-\ref{scales}. It can be seen that for super-Planckian $\phiast$, the predictions match with that of the large-field chaotic models. Interestingly for $n=4$, with the decreasing $\phi_*$ the  value of $\lambda$ increases, which could be useful in the context of pure Higgs inflation without the non-minimal curvature coupling. Another important point to notice that for $\phi_*<\mbox{M}_p$, all the relevant scales and importantly the field excursion became sub-Planckian, which is one of the important criteria for an effective field theory to be valid.

\subsection{End of inflation and general equation of state}
In this section we will be interested in the dynamics of the inflaton field
after the inflation. During this phase the inflaton field oscillates coherently around the minimum
of the potential. At the beginning the oscillation dynamics will naturally be dependent upon 
the inflation scale $\phi_*$ because of the large amplitude. This is the stage during which non-perturbative
particle production will be effective. Therefore, resonant particle production will take place and
conversion of energy from the inflaton to matter particles will be highly efficient. This phenomena 
is usually known as pre-heating of the universe. In this section we will discuss about
the late time behaviour of the inflaton, specifically focusing on the dynamics of the
energy density of the inflaton field. After many oscillations, 
when the amplitude of the inflaton decreases below $\phi_*$, the dynamics will be
controlled by usual power law potential. As we have emphasized the coherent oscillation is 
very important in standard treatment of reheating. 
For any model of inflation this is thought to be an important criterion to have successful reheating.  
In this section, we will first discuss the evolution of inflaton and its energy density. At late time the potential can naturally be approximated as
\be
\lim_{\frac{\phi}{\phi_*} <1}V_{\rm min}(\phi) = \lambda m^{4-n} \phi^n .
\ee

\begin{table}[t!]
\begin{tabular}{|p{0.6cm}|p{2.0cm}|p{2.5cm}|p{2.5cm}| }
\hline
$n$& $ \textit{w}   =\frac{n-2}{n+2} $ & $p = 3(1 + \textit{w})$& $p$ from fitting\\
\hline
\hline
$2$ & 0 & 3 & 3.12 \\
\hline
4 & $\frac{1}{3}$ & 4 & 3.99 \\
\hline
6 & $\frac{1}{2}$ & 4.5 & 4.56 \\
\hline
8 & $\frac{3}{5}$ & 4.8 & 4.83 \\
\hline
\end{tabular}
\caption{The variation of inflation energy density with scale factor for various potential}
\label{T-rhoVa}
\end{table}

In cosmology for any dynamical field such as inflaton, one usually defines the equation of state parameter $\textit{w}$.
For the oscillating inflaton, when the time scale of oscillation about the minimum of a potential is small enough compared
to the background expansion time scale, by using virial theorem effective equation of state for a potential of the 
form $V(\phi) \propto \phi^n$ can be expressed as\cite{Mukhanov:2005sc}
 \be
 \textit{w} \equiv \frac{P_{\phi}}{\rho_{\phi}} \simeq \frac{\langle \phi V'(\phi) \rangle - \langle 2 V\rangle}{\langle \phi V'(\phi)\rangle + \langle 2 V\rangle}=\frac{n-2}{n+2} .
 \label{w-n}
 \ee
Therefore, in an expanding background, the evolution of energy density $\rho_{\phi}$ of the inflaton averaged over many oscillation
will follow,
\bea
\dot{\rho}_{\phi} + 3 H (1+\textit{w}) \rho_{\phi} = 0 .
\eea
At late time we relate the energy density$(\rho_{\phi})$ of the universe
(assuming that the universe is dominated by a single component) and the scale factor $(a)$ as
\be
 \rho_{\phi} \propto a^{-3(1+\textit{w})} = a^{-p} .
 \label{rho-n}
 \ee
In the table-\ref{T-rhoVa}, we provide some theoretical as well as numerically fitting values corresponding to 
the equation of state parameter $\textit{w}$ of the inflaton and the power law evolution of the energy 
density namely the value of $p$.   
 

Before we come to reheating analysis, in the next section we describe our model potential originating from the supergravity in a specific limit.        

\section{\label{sugra}Supergravity realization of our model potential
	}
As we have already mentioned, in this section we will construct our non-polynomial potential from supergravity generalizing the construction proposed in \cite{Dimopoulos:2016zhy} for $n=2$. 
Let us first briefly review the scalar field inflation from supergravity\cite{Mazumdar:2010sa,Yamaguchi:2011kg,Nakayama:2016eqv}. For supergravity model of inflation, the usual approach is to consider the K\"ahler potential $K(\Phi_i, \Phi_{\bar{i}}^{\ast})$, the superpotential $W(\Phi_i)$ in terms of scalar superfields $\{\Phi_i\}$ with the associated Lagrangian in the Einstein frame
\begin{equation}
 \mathcal{L} = K_{i\bar{j}} (\partial^{\mu}\Phi_i)(\partial_{\mu}\Phi_{\bar{j}}^{\ast}) - V.
\end{equation}
However for our present purpose, we also need to consider gauge kinetic function $f(\Phi_i)$, which couples with the gauge field kinetic term. Considering all those terms in the supergravity Lagrangian, one obtains $F$-term and $D$-term
potential for the inflaton and other associated moduli fields as,  
 \begin{eqnarray}
  V   &=& V_F + V_D,\\
  V_F &=& e^{K/{\rm M_p^2}} \left[ K^{i\bar{j}}(D_i W)(D_{\bar{j}}\bar{W}) -3\frac{|W|^2}{\rm M_p^2}\right],\\
  V_D &=& \frac{g^2}{2}\mathfrak{R}(f)^{-1}(iK_i X_i)^2,
 \end{eqnarray}
where $K^{i\bar{j}}( = K_{i\bar{j}}^{-1})$ is the inverse  K\"ahler metric $K_{i\bar{j}}$. $D_i W = W_i + K_i W/{\rm M_p^2}$ is the K\"ahler covariant derivative, and $X_i$  is the Killing vector of the K\"ahler manifold, and $g$ is the gauge coupling constant. For a linearly transforming field under  $U(1)$ symmetry we have $X_i = i q_i \Phi_i$, where $q_i$ is the $U(1)$ charge of $\Phi_i$. The D-term potential in this case reduces to
\begin{equation}
 V_D = \frac{1}{2\mathfrak{R}(f)}\left( \sum_i q_i K_i \Phi_i  + \xi_i \right)^2 ,
\end{equation}
where $\xi_i$ is the known as Fayel-Iliopoulos (FI) term which is non-zero only when the gauge symmetry is Abelian. 

\subsection{The power-law plateau potential form Supergravity}
Let us consider a particular form of the aforementioned superpotential with two chiral superfields $S$ and $\{\Phi_i\}$ as
\begin{equation}
 W(S, \Phi_i) = \frac{\phiast^{3-n}S^n}{n} F(\Phi_i) .
\end{equation}
Where, $\phiast$ mass scale and $F$ is a dimensionless holomorphic function of the superfields $\Phi_i$. The canonical K\"ahler potential is taken to be
\begin{equation}
 K = |S|^2 + \sum_i |\Phi_i|^2 .
 \end{equation}
With the above ingredients we can straightforwardly compute the  $F$-term potential as,
\begin{align}
 V_F = \phiast^{2(3-n)}e^{K/{\rm M_p^2}} \left\{ |F|^2|S|^{2(n-1)}\left(1 + \frac{|S|^2}{n{\rm M_p^2}}\right) + \frac{|S|^{2n}}{n^2}\left| \frac{\partial F}{\partial \Phi_i} + \frac{\Phi_i^{\ast}F}{\rm M_p^2} \right|^2 -3\frac{|W|^2}{\rm M_p^2}  \right\} .
\end{align}
For our present purpose, let us consider $n=2$ and two superfields with the following form of $F$,
\begin{align}
W(S,\Phi_1, \Phi_2) =& \frac{\phiast S^2}{2}F(\Phi_1, \Phi_2)\\
\nonumber
\text{with,\quad} F(\Phi_1, \Phi_2) =& F_1(\Phi_1) - F_2(\Phi_2)
\end{align}
Further we assume the functional form of the holomorphic functions $F_1$ and $F_2$ to be same, i.e., $F_1 \equiv F_2$. The $D$-term potential for a suitable gauge coupling function and charge we will take from the reference \cite{delaMacorra:1995qh,Dimopoulos:2016zhy}, and the expression is given as
\begin{equation}
 V_D = \frac{1}{2} \left( |S|^2 - \sqrt{2}M^2 \right)^2
\end{equation}
where $M$ is another scale which is associated with the FI term. As mentioned before the D-term potential is generated considering anomalous $U(1)$ symmetry\cite{Binetruy:1996xj,Halyo:1996pp}. These type of symmetries usually appears in the realm of string theories\cite{Dine:1987xk,Atick:1987gy,Dine:1987gj}. The anomaly cancellation requires the Green-Schwarz (GS) mechanism\cite{Green:1984sg} which sets the value of the FI term as
\begin{equation}
\xi_{\rm GS} = \frac{Tr[Q_A]}{192\pi}g^2 M^2 ,
\end{equation}
where, $Q_A$ is the charge of the fields under the anomalous $U(1)$ gague and trace is taken over fields. 
Now let us assign the charges of the fields following the discussion in\cite{Halyo:1996pp}. The charge of the fields $\Phi_1$ and $\Phi_2$ (whose modulus will be identified as the inflaton) are assumed to be zero. The charge of $S$ will be non-zero as long as it is opposite of ${\rm Tr}[Q_A]$, and is taken to be $-1$.(This will make it necessary to consider the existence of another field charged under the anomalous $U(1)$ with zero VEV.)

\begin{figure}[t]
 	\includegraphics[scale=0.6]{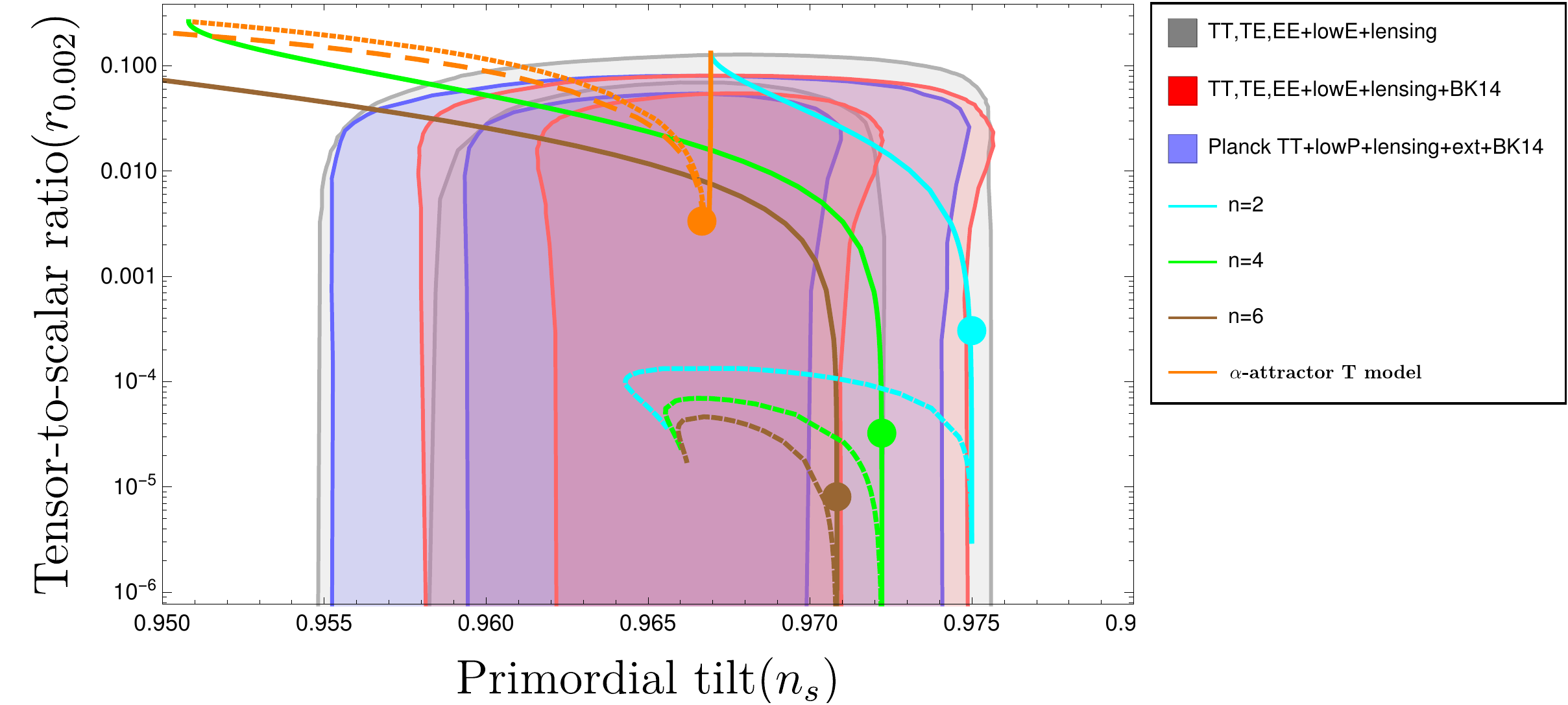}
 	\caption{\scriptsize The $n_s$-$r$ plot ffor $60$-efolding number of the two models plotted on the Planck data\cite{Akrami:2018odb}. The solid lines corresponds to the minimal model while the dotted lines are for the SUGRA model. We have varied the values of $\phiast$ in the range $(10^{-3}\mathrm{M_p},10^2\mathrm{M_p})$. The dots are the theoretical prediction from Eq.(\ref{nsrvsN}) for $\phiast=0.1~{\rm M_p}$. The orange lines correspond to the $\alpha$-attractor T model with solid line being $n=2$, and densely dashed and dashed lines for $n=4$, and $n=6$ respectively. We have varied $\alpha=1$ to $\alpha=10^4$. The orange dot corresponds to the Starobinsky model($\alpha=1$) for the same efolding number.}
 	\label{pplanck}
 \end{figure}

With the above considerations, the total potential for inflaton $\phi$ and the moduli $S$ will be $V_T = V_D + V_F$. Given the simple form of the superpotential, the minimization along ${\Phi_i}$ leads to the condition $\Phi_1=\Phi_2$. Now assuming a suitable R-symmetry and considering a particular direction in field space as $|\Phi_1|=|\Phi_2| =\phi$, the total scalar potential for sub-Planckian value of $|S|$ simplifies to 
\begin{align}
 V_T = e^{\frac{\phi}{\rm M_p^2}} \phiast^2 |S|^4  \left|F'(\phi)\right|^2 + \frac{1}{2} \left( |S|^2 - \sqrt{2}M^2 \right)^2
\end{align}
 \begin{figure}[!t]
	\includegraphics[scale=0.5]{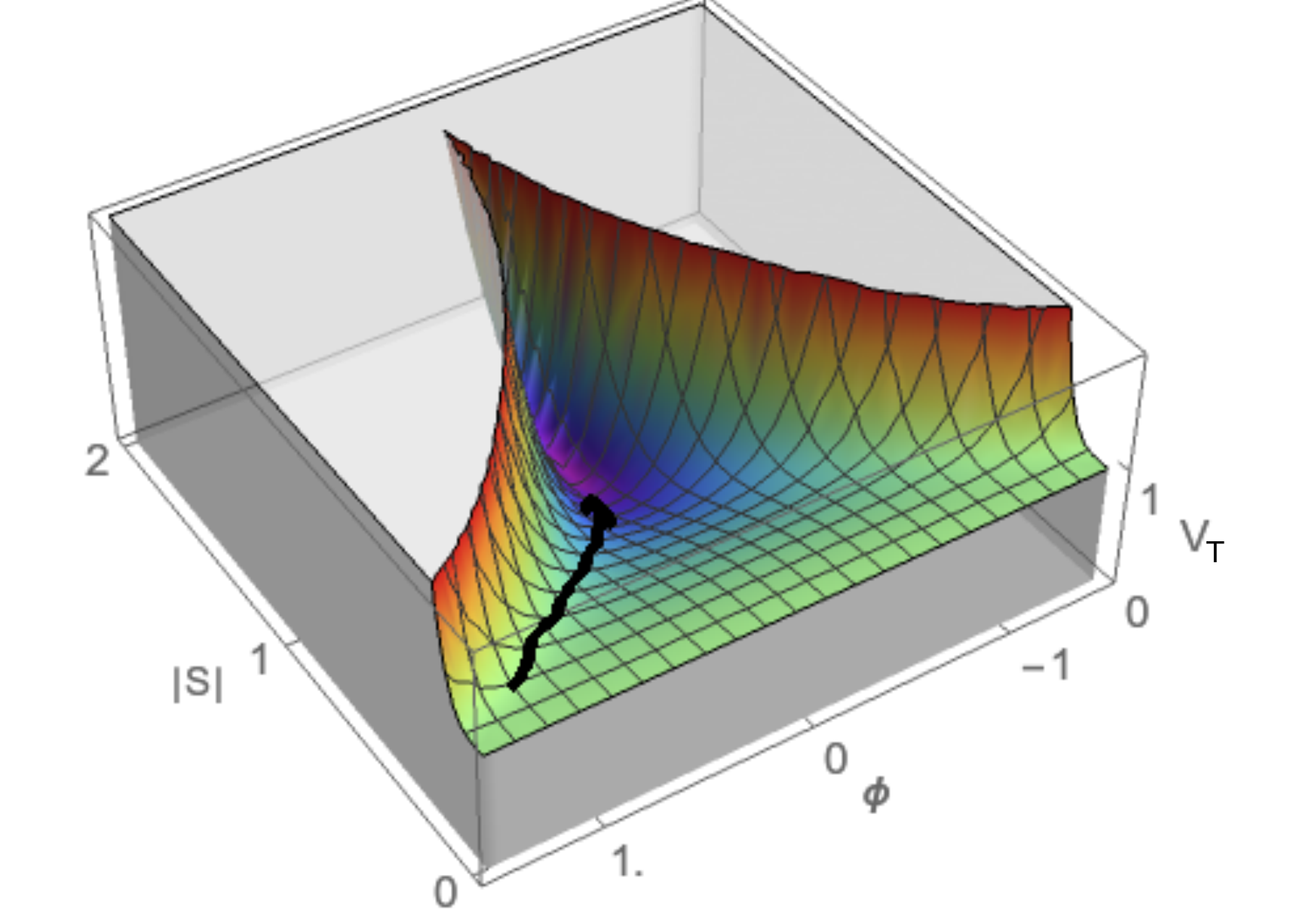}
	\caption{ Total supergravity potential $V_T$ has been plotted. All the field are plotted in unit of Planck unit with $\phi_*= M = {\rm M_p}$. The black curve corresponds to a possible inflationary trajectory around the minimum direction of $|S|$ which is the scalar field potential $V_{\rm sugra}$.}
	\label{sugra3d}
\end{figure}
In fig.\ref{sugra3d} we have given simple illustration of our supergravity potential in $(|S|,\phi)$ space and the inflationary trajectory. In order to obtain an effective  potential in term of inflaton $\phi$, we minimize $V_T$ along the $S$ direction and plug it back into the total potential which takes the following simple form
\begin{equation}
 V_{T}(\phi) = M^4\left[\frac{ e^{\left(\frac{\phi}{\rm M_p}\right)^2} \phiast^2 \left|F'(\phi)\right|^2}{1 + e^{\left(\frac{\phi}{\rm M_p}\right)^2} \phiast^2 \left|F'(\phi)\right|^2} \right]
 \label{eq:sugrapot0}
\end{equation}
It is now easy to check that for $F(\varphi)\propto (\phi/\phiast)^p$ ($\{ p \in \mathbb Z \mid p > 1 \}$ ) we get a potential of the form
\begin{equation}
 V_{\rm sugra}(\phi) = \frac{m^{4-n}\phi^n}{{\rm exp}(-\frac{\phi^2}{2{\rm M_p^2}}) + \left( \frac{\phi}{\phi_{\ast}}\right)^n }
 \label{eq:sugrapot}
\end{equation}
Where $n=2(p-1)$ and the scale $m$ is defined by combining the scale $M$ with $\phiast$ as $m=(M^4\phiast^{-n})^{1/(4-n)}$. In the limit $\phi_* < {\rm M_p}$, this potential reduces to the minimal plateau model given as,
\begin{equation}
 V_{\rm min}(\phi) = \frac{m^{4-n}\phi^n}{1 + \left( \frac{\phi}{\phi_{\ast}}\right)^n }
\end{equation}
\begin{figure}[t!]
	\centering
	\subfigure[]{\includegraphics[scale=0.4]{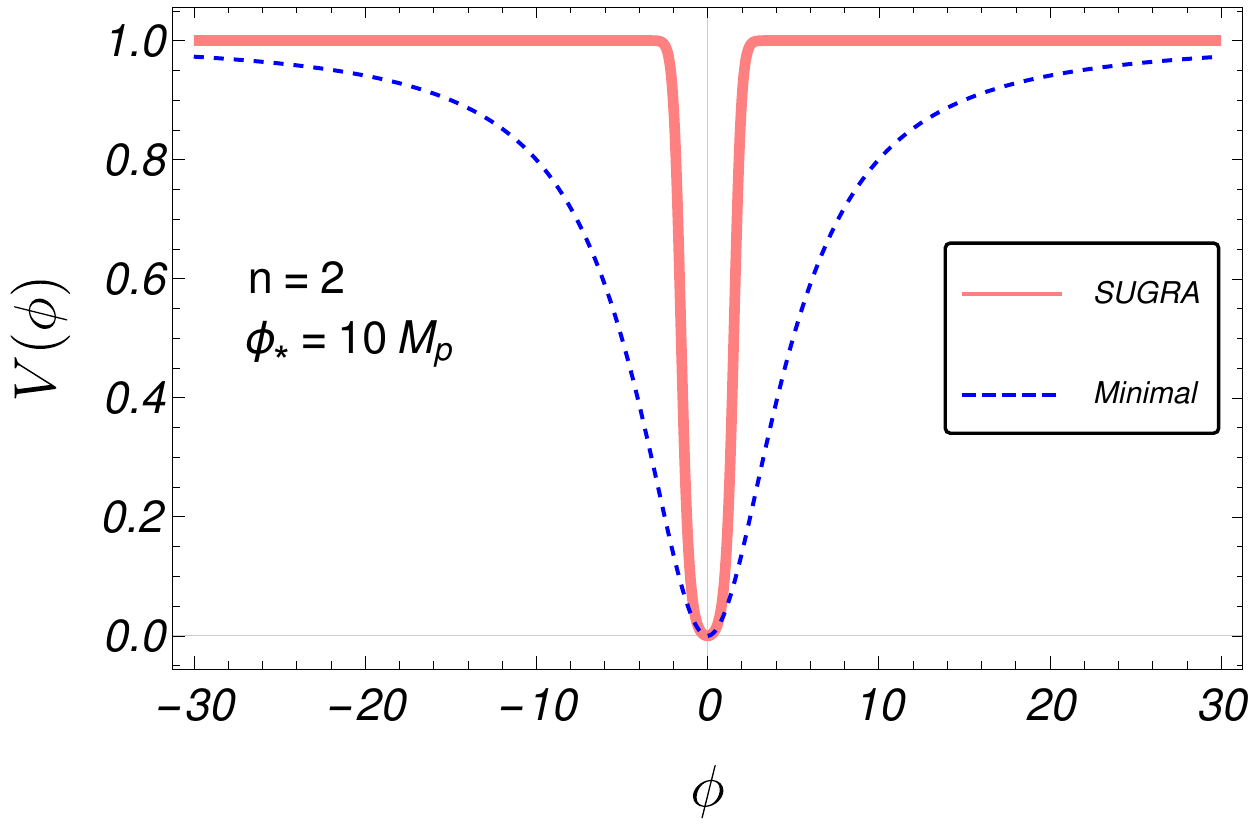}}
	\subfigure[]{\includegraphics[scale=0.4]{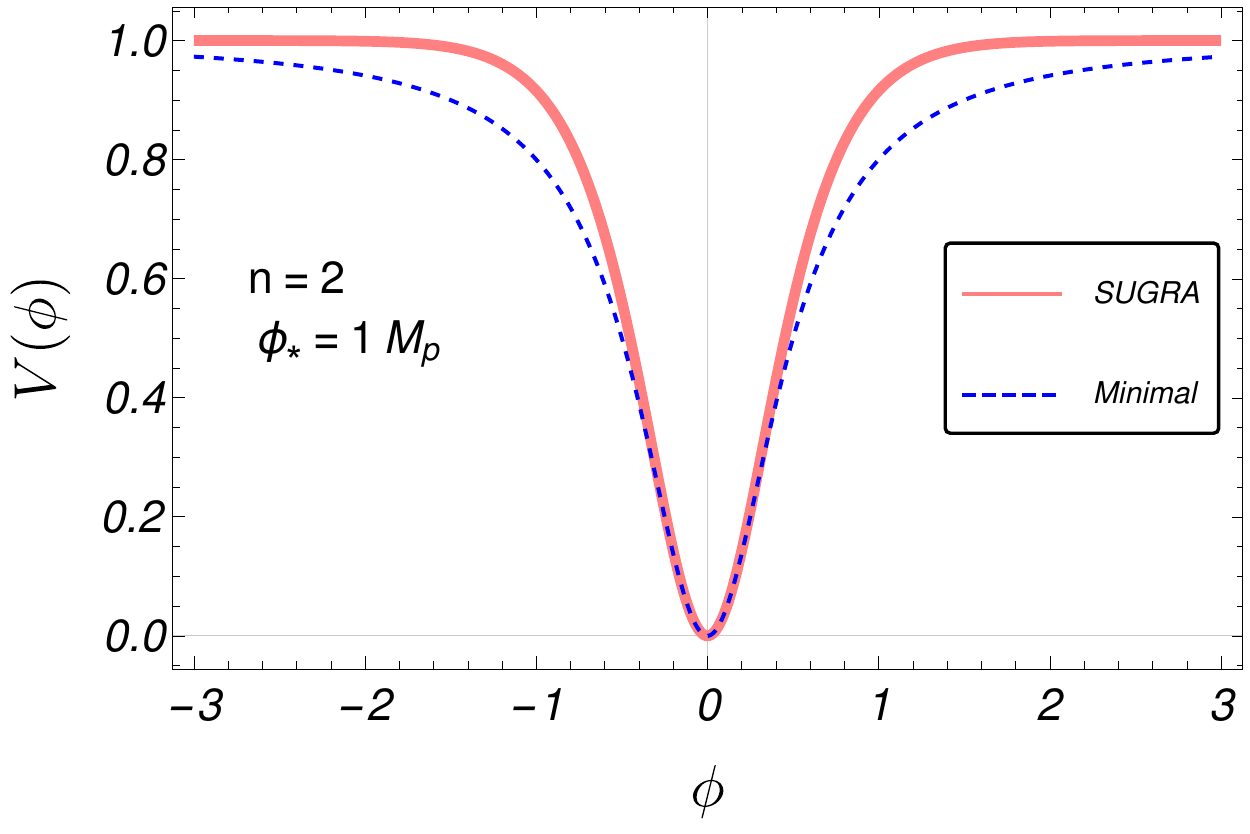}}
	\subfigure[]{\includegraphics[scale=0.4]{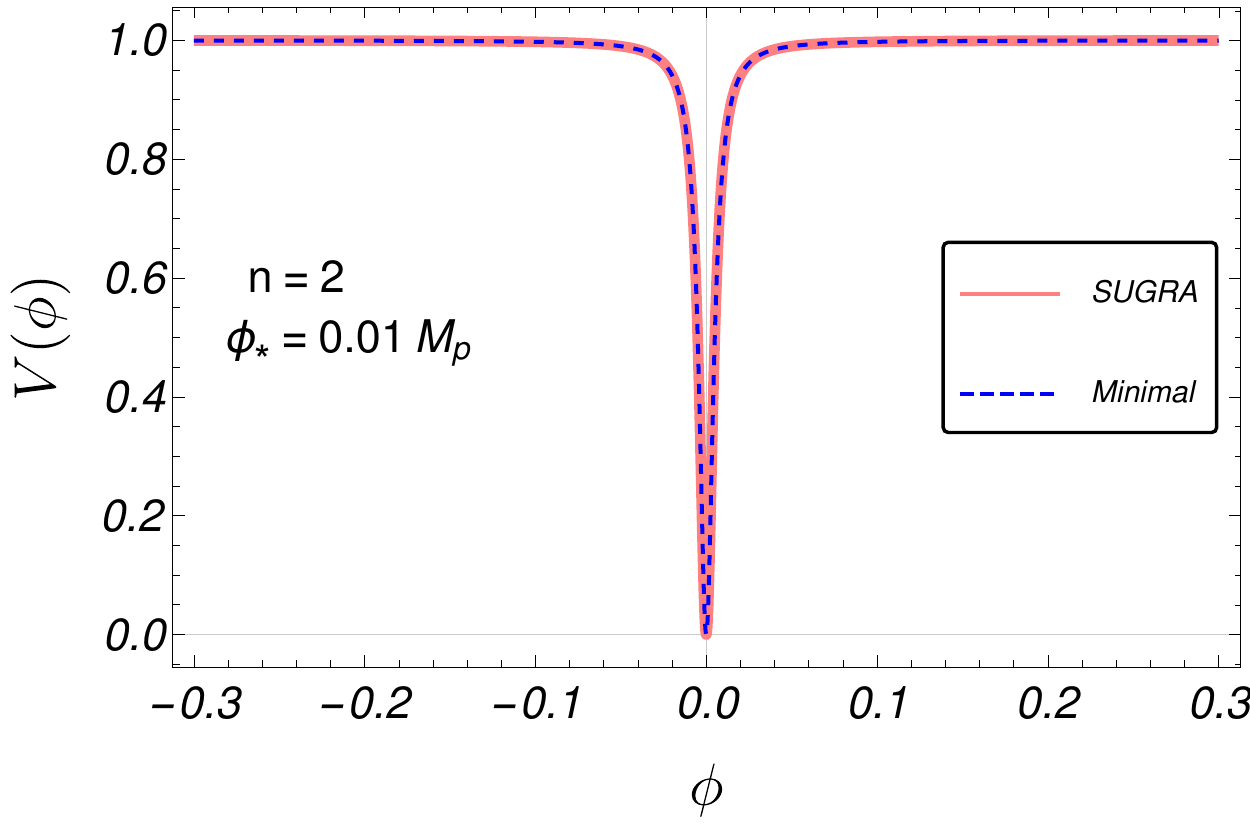}}
	\caption{\scriptsize Comparison of the shape of the two potentials for three different values the scale $\phiast$. It can be seen that for subplanckian values of $\phiast$ the two potentials are identical.}
	\label{fig:pots1}
\end{figure}
\begin{figure}[!]
	\centering
	\subfigure[]{\includegraphics[scale=0.6]{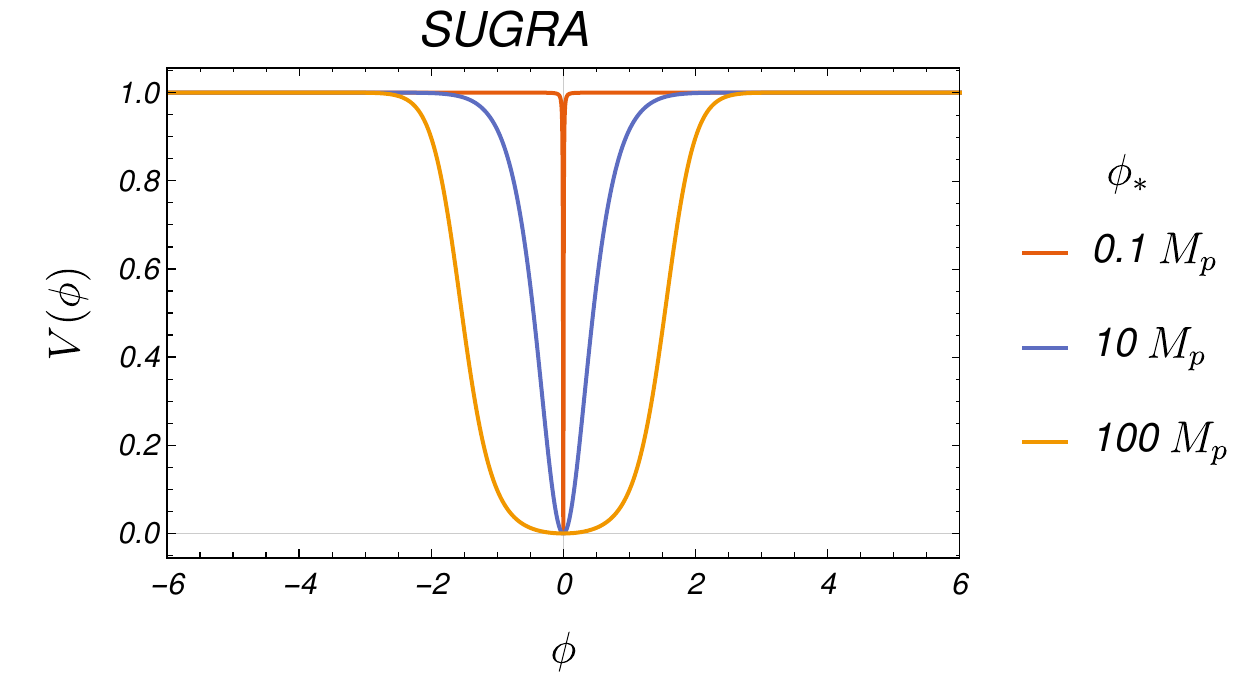}}
	\subfigure[]{\includegraphics[scale=0.6]{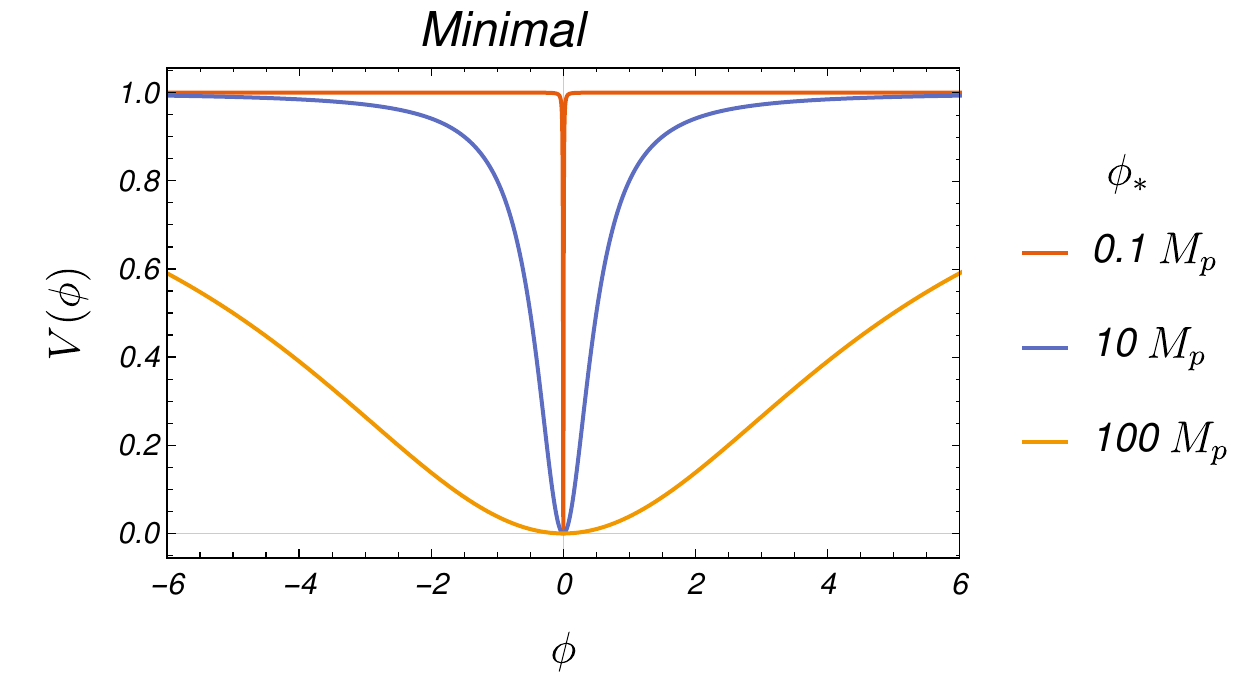}}
	\caption{\scriptsize The variation of the shape of the two potentials with scale $\phiast$. The width of the potential vary slowly with $\phiast$ for the plateau potential compared to the minimal potential.}
	\label{fig:pots2}
\end{figure} 
In our subsequent discussions, we will try to compare both aforementioned potentials as a separate entity. In fig.(\ref{fig:pots1}), we plotted them for illustrations corresponding to different values of the parameters. It is apparent that the supergravity potential $V_{\rm sugra}$ and our minimal plateau potential $V_{\rm min}$ are identical for the sub-Planckian values of the scale $\phiast$. Therefore, for our subsequent analysis, we will mostly concentrate in the sub-Planckian region of $\phi_*$. At this point we also would like to point out that we can arrive at the same form of our minimal potential if we start from the non-minimal scalar filed theory as discussed in the appendix. 
\begin{figure}[!]
	\centering
	\subfigure[]{\includegraphics[scale=0.6]{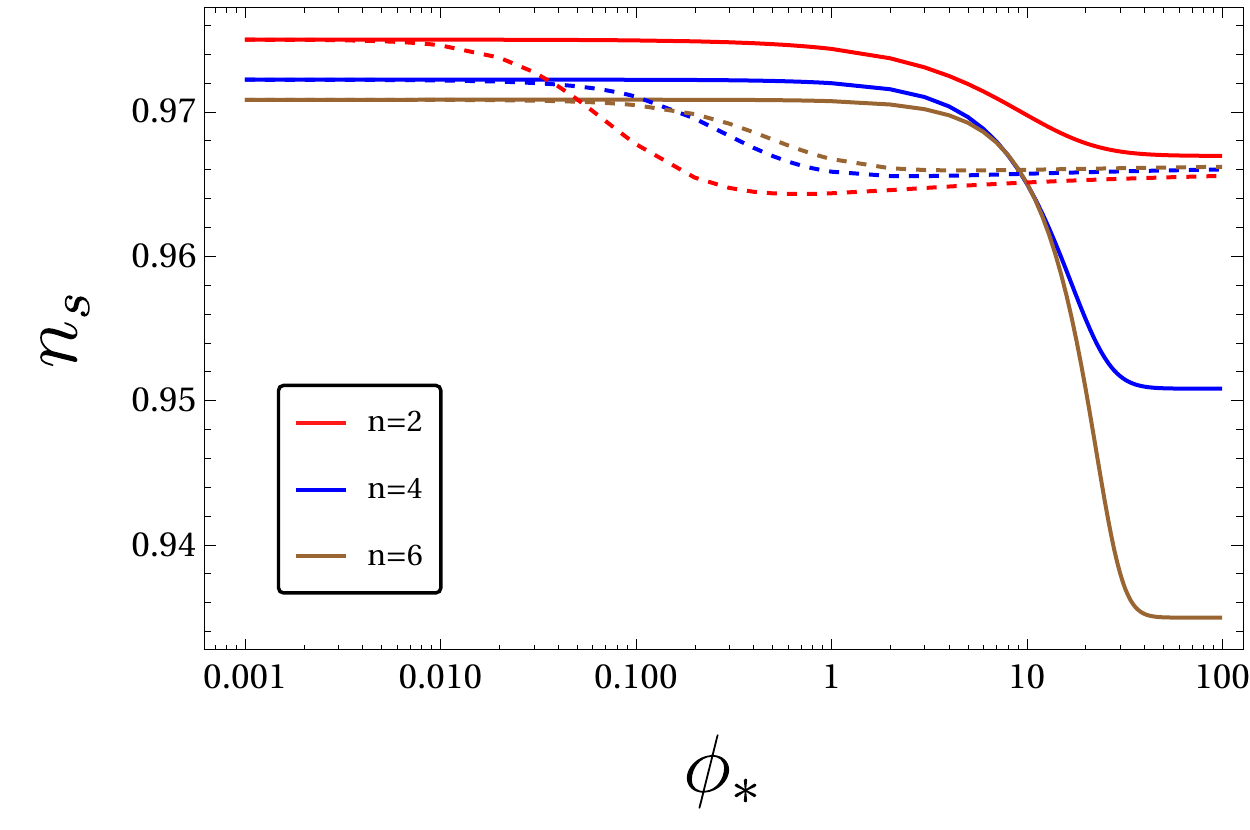}}
	\subfigure[]{\includegraphics[scale=0.6]{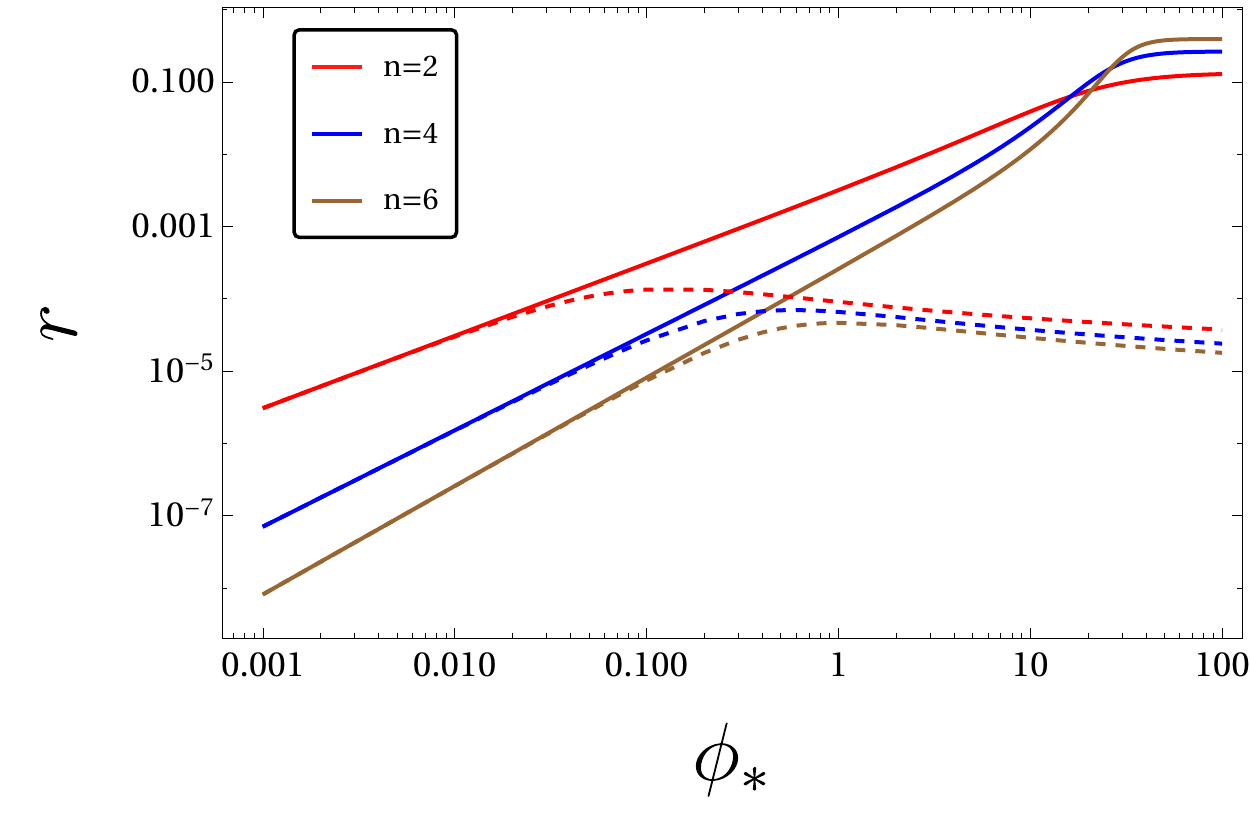}}
	\caption{\scriptsize variation of the spectral index $(n_s)$ and scalar-to-tensor ratio with the scale $\phiast$ for the two potentials. The solid lines corresponds to the minimal model while the dotted lines are for the SUGRA model. For smaller values of $\phiast$ the two model predictions are identical.}
	\label{fig:mnsr}
\end{figure}
Nevertheless in Fig.(\ref{fig:pots2}) we illustrate the change of the shape of the two potentials with the scale $\phiast$. The $V_{\rm sugra}$ is extremely flat even for large value $\phiast$, while the minimal potential reduces to the simple power law potentials. This fact significantly controls the prediction of inflationary observables for both the models. In Fig.(\ref{fig:mnsr}), we have plotted the dependence of the $n_s$ and $r$ on the scale $\phiast$. We clearly notice that the  predictions of both the models are identical as we go towards sub-Planckian value of $\phi_*$. In the Fig.(\ref{pplanck}), we have plotted predictions of our different models in $n_s$-$r$ plane. For explicit comparison, we have also plotted the predictions of the $\alpha$ attractor $T$ model($V=\Lambda^4[1-exp(-\sqrt{2/3\alpha}~\phi/{\rm M_p})]^{n}$) for different $n$ with increasing values of the parameter $\alpha$ with $(n=2,\alpha=1)$ being the Starobinsky model. Interestingly, the super-gravity model turned out to be well within the $2\sigma$ region of $n_s$ at all scales. Even at large field value of the inflaton the prediction of $r$ is always small. On the other hand as discussed before, our minimal potential mimics the usual chaotic inflation at large field value. Another interesting point that is worth mentioning is the absence of attarctor behavior in terms of cosmological predictions for the minimal plateau models. As can be seen from the Fig.(\ref{pplanck}), for our supergravity potential and the $\alpha$-attractor potential the value of $(n_s,r)$ going towards their respective unique attractor value with increasing $\phi_*$ and $\alpha$ respectively irrespective of any other parameter values. For $\alpha$-attractor this unique value is that of Starobinsky model with $(n_s=0.9667, r=3\times10^{-3})$, and for our supergravity potential the unique attractor value turned out to be $(n_s=0.9667, r=1\times10^{-5})$. It would be interesting to find that specific theory for this particular prediction. For the $\alpha$-attractor models, these attractor behavior is due to the pole in the kinetic term\cite{Galante:2014ifa}. As we have considered only the canonical kinetic term in our Lagrangian, the attractor behavior is absent for our minimal plateau models(see also appendix-\ref{appendix:appA} for the region when this approximation is valid). However, the emergence of attractor behavior for the original supergravity potential given in (\ref{eq:sugrapot0}) is indeed an interesting phenomena. The detailed theoretical implications of this supergravity potential will be important which we leave for future study. In our subsequent discussion we will only concentrate on the power-law minimal plateau potentials.
In the next section, will be considering the equation of state parameters and study their role in the subsequent cosmological evolution. We will first discuss about the constraint on reheating phenomena by taking the model
independent approach, where explicit dynamics during reheating phase will not be considered. 

\section{\label{reheatingprediction}Model independent constraints from reheating predictions}
  After inflation, reheating is the most important phase, where, all the visible matter energy
  will be pumped in. In this section, we will try to constrain our model parameters 
  without any specific mechanism of reheating. The background
evolution of cosmological scales from inflation to the present day and the conservation of entropy density
provide us important constraints on reheating as well as our model parameters.  
Reheating is the supposed to be the integral part of the inflationary paradigm. 
However, because of the single observable universe, it 
is very difficult to understand this process by the present day cosmological observation.
Thermalization process erases all the information about the initial conditions which is the most
important part of this phase. To understand this phase an indirect attempt has been made in the recent past \cite{liddle,kamionkowski,cook} 
through the evolution of cosmological scales and the entropy density. The dynamics is parameterized 
by three independent parameters, reheating temperature $(T_{re})$, equation of state $(w_{re})$, and e-folding 
number $(N_{re})$. In this section we follow the reference \cite{debuGB}, and consider two stage  reheating process generalizing the formalism of \cite{liddle}. 
Because of two stage reheating process, the suitable reheating parameters 
are as follows, $(N_{re}=N^1_{re}+N^2_{re},T_{re},w^1_{re},w^2_{re})$. 
Where, $N^1_{re}, N^2_{re}$ are e-folding number during the first and second stage of the reheating 
phase with the equation of state $w^1_{re}, w^2_{re}$ respectively. 
At the initial stage the oscillating inflaton will be the dominant component, and
at the end radiation must be the dominant component. Therefore, one sharp contrast between our present analysis with that of the well known reheating constraint analysis \cite{kamionkowski} is that our reheating equation of state $w_{re}$ is no  longer a free parameter.  We rather consider the following particular case, 
 \bea
w^1_{re} = \frac{n-2}{n+2} ~~;~~~ w^2_{re} = \frac 1 3 .
\label{eqparameter}
\eea
 We also assumed the change of reheating phase from the first to the 
second stage as instantaneous.


A particular scale $k$ going out of the horizon during inflation will re-enter the horizon during
usual cosmological evolution. This fact will provide us an important 
relation among different phases of expansion parameterizing by e-folding number as follows 
\bea
\ln{\left(\frac k {a_0 H_0}\right)} = \ln{\left(\frac {a_k H_k}{a_0 H_0}\right)} =-N_k -\sum_{i=1}^{2}N^i_{re} -
\ln{\left(\frac {a_{re} H_k}{a_0 H_0}\right)},
\label{scalek}
\eea  
In the above expressions, use has been made of $k = a_0 H_0 = a_k H_k$. 
Where, $(a_{re}, a_0)$ are the cosmological scale factor at the end of the reheating phase and the present time respectively.
$(N_k,H_k)$ are the e-folding number and the Hubble parameter respectively for a particular scale $k$ 
which exits the horizon during inflation. Therefore, following mathematical expressions will be used in the final numerical
calculation, 
\begin{eqnarray}
H_k &=&  \sqrt{\frac{V(\phi_k)}{3 {\rm M_p^2}}} = \begin{cases} \left(\frac{\lambda \phi_*^n}{3 {\rm M_p^2}}\right)^{\frac 1 2} 
\frac{m^{\frac{4-n}{2}} \tilde{\phi}_k^{\frac {n}{2}}}
{\left(1+\tilde{\phi}_k^n\right)^{\frac 1 2}} \\ 
 \left(\frac{\lambda \phi_*^n}{\rm 3 M_p^2}\right)^{\frac 1 2} \frac{m^{\frac{4-n}{2}} 
 \tilde{\phi}_k^{\frac{n}{2}}}{\left(1+\tilde{\phi}_k^2\right)^{\frac{n}{4}}},
 \end{cases}
\\
N_k &=&  \frac 1 {\rm M_p} 
\int_{\phi_{end}}^{\phi_{k}} \frac {1} {\sqrt{2 \epsilon}} d\phi \simeq
\begin{cases}
          \frac{\phi_*^2}{n {\rm M_p^2}}\left[\frac{1}{(n+2)}\tilde{\phi}_k^{(n+2)} + \frac{1}{2} \tilde{\phi}_k^2\right] \\
         \frac{\phi_*^2}{n {\rm M_p^2}} \left[\frac{1}{4} \tilde{\phi}_k^4 + \frac{1}{2} \tilde{\phi}_k^2\right] ,
       \end{cases}
       \label{HandN}
\end{eqnarray}
\begin{figure}[t!]
\begin{center}
  \includegraphics[width=006.0cm,height=04.0cm]{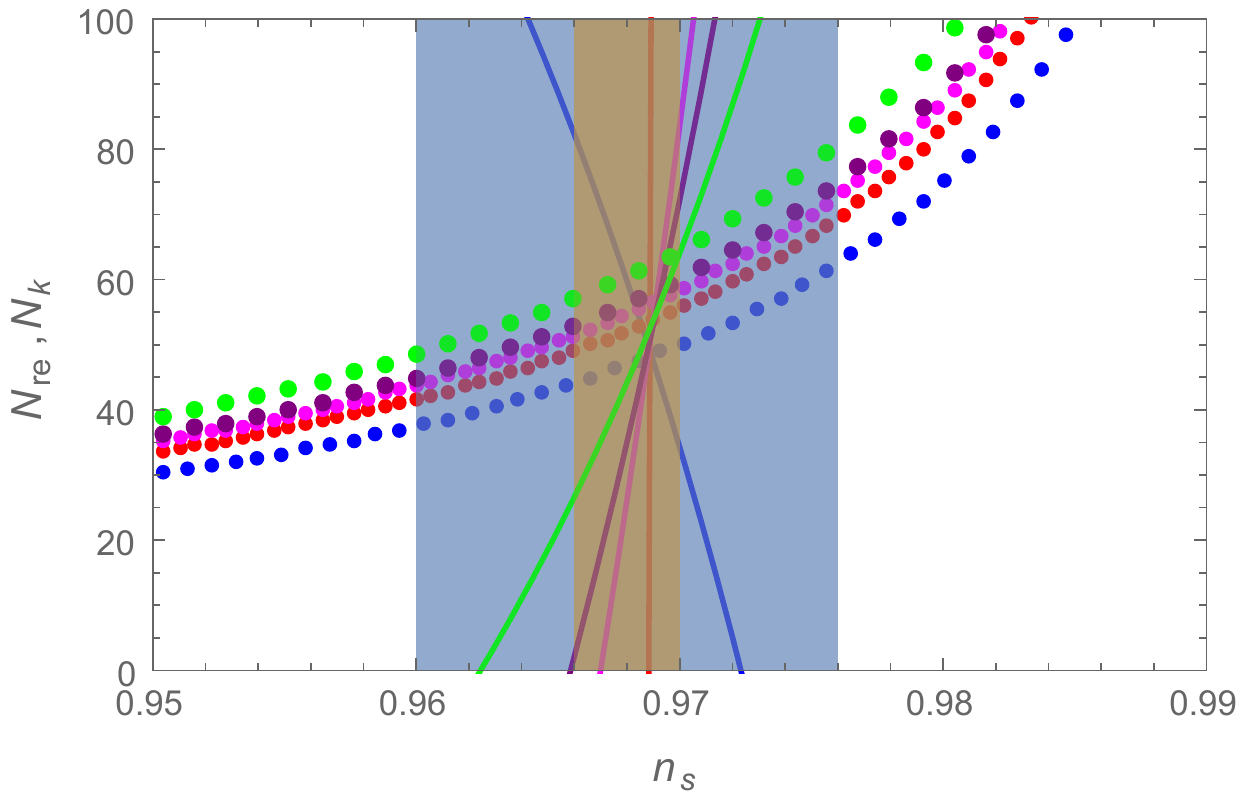}
  \includegraphics[width=006.0cm,height=04.0cm]{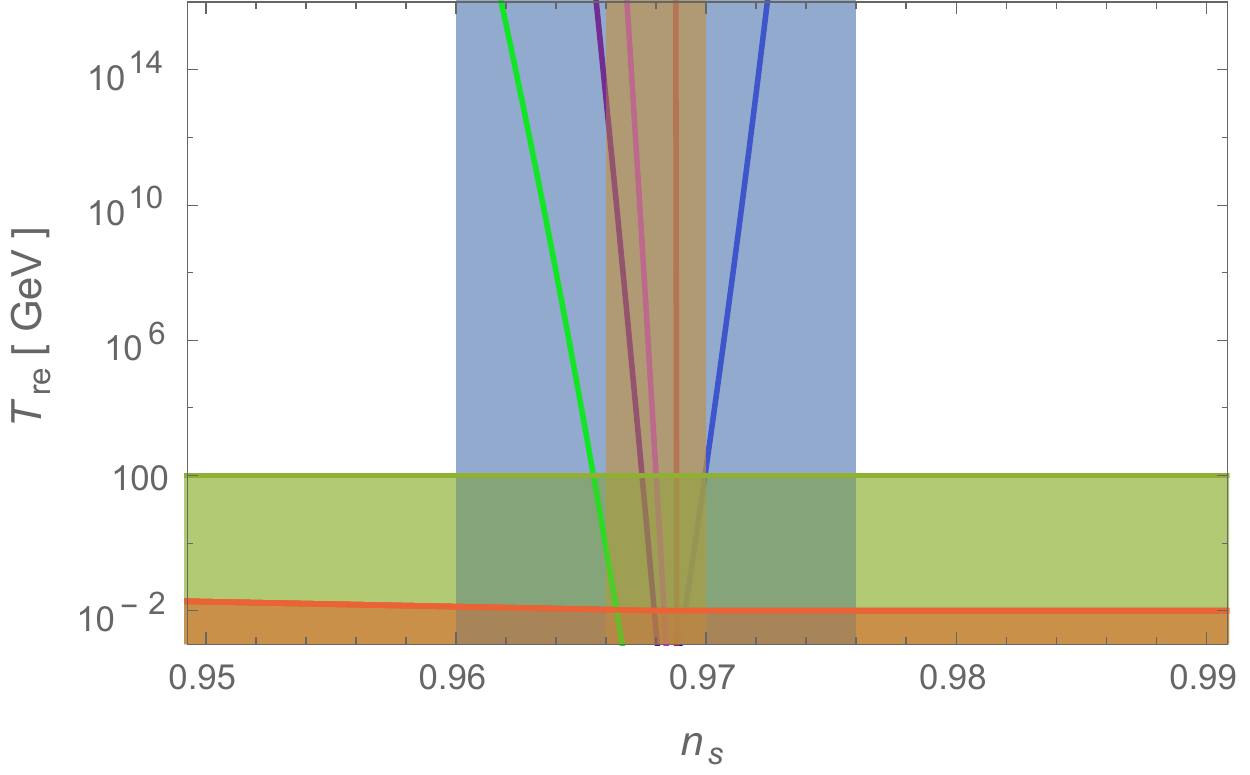}
 \caption{\scriptsize Variation of $(N_{re}(\mbox{solid}), N_{k}(\mbox{dotted}),T_{re})$ as a function of $n_s$ have been plotted for 
$\phi_* = 0.01 {\rm M_p}$. This is the plot for our model potential. (Blue, red, magenta, brown, green) curves correspond to $n=(2,4,6,8,30)$. 
Each curve corresponds to a specific set of equation of state parameters $(w^1_{re},w^2_{re}) =((n-2)/(n+2),1/3)$ during reheating.
We also consider $N^1_{re}=N^2_{re}$. The light blue shaded region 
corresponds to the $1 \sigma$ bounds on $n_s$ from Planck. The brown shaded region corresponds to the $1 \sigma$
bounds of a further CMB experiment with sensitivity $\pm 10^{-3}$ \cite{limit1,limit2}, using the same central $n_s$ value as
Planck. Temperatures below the horizontal red line is ruled out by BBN. The deep green shaded region is below the 
electroweak scale, assumed 100 GeV for reference.} 
\label{tre1}
\end{center}
\end{figure}
where, $\phi_k$ and $\phi_{end}$ are the inflaton field values corresponding to a particular scale $k$ crossing the inflationary horizon,
and at the end of inflation respectively.
In the above expressions, we have ignored the contribution coming from the inflaton field value $\phi_{end}$.
It is important to note that, in principle we can write the field value at a particular scale $k$
in terms of $n_s, r$, by inverting those relations. Because of non-linear form, we will numerically solve those.
The above unknown efolding numbers during reheating will certainly be dependent upon the energy densities 
$(\rho_{end},\rho_{re})$, at the end of inflaton (beginning of reheating phase) and at the end of the 
reheating phase( beginning of the standard radiation dominated phase);  
\bea
\ln\left(\frac {\rho_{end}}{\rho_{re}}\right) = 3(1+ w^1_{re}) N^1_{re} + 3(1+ w^2_{re}) N^2_{re}= 3\sum_{i=1}^{2} (1+ w^i_{re}) N^i_{re}.
\label{rho2}
 \eea
Above two Eqs.(\ref{scalek}) and (\ref{rho2}), can be easily generalized for multi-stage reheating with different 
equation of state parameters. 
As has been mentioned, after the end of reheating standard evolution of our universe is precisely known in terms of
energy density and the equilibrium temperature of the relativistic degrees of freedom such as photon and
the neutrinos. Therefore, the equilibrium temperature after the end of reheating phase, $T_{re}$, is related
to temperature $(T_0, T_{\nu 0})$ of the CMB photon and neutrino background at the present day respectively, as follows
\bea \label{entropy}
g_{re} T_{re}^3 = \left(\frac {a_0}{a_{re}}\right)^3\left( 2 T_0^3 + 6 \frac 7 8 T_{\nu 0}^3\right).
\eea
\begin{figure}[t]
	\begin{center}
		\includegraphics[width=8.0cm,height=02.8cm]{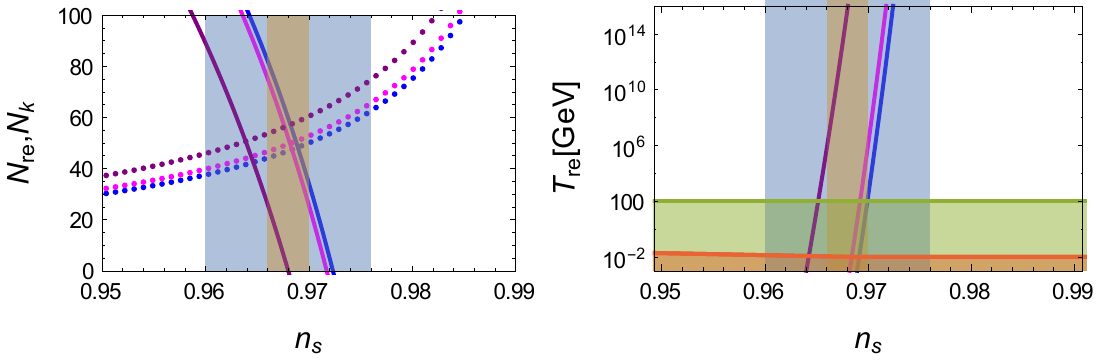}
		\includegraphics[width=8.0cm,height=02.8cm]{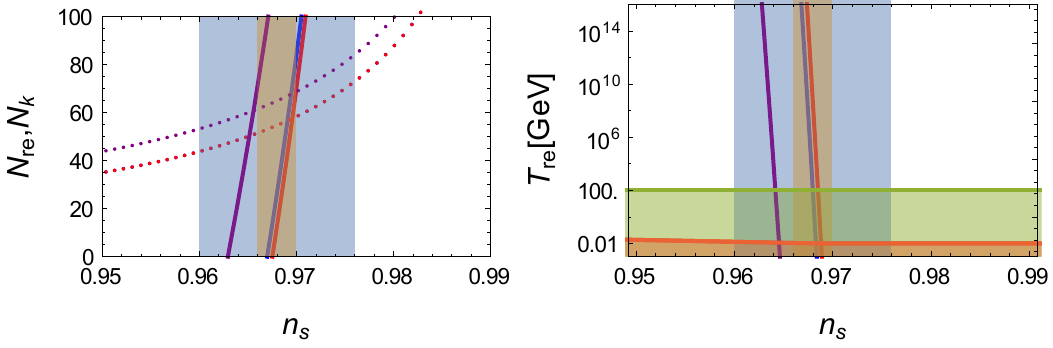}
		\caption{\scriptsize Variations of $(N_{re}(\mbox{solid}), N_{k}(\mbox{dotted}),T_{re})$ as a function of $n_s$ have been plotted for three different values of $\phi_*$. (Blue, magenta, purple) curves are for $\phi_* = (0.01, 0.1,10) {\rm M_p}$ respectively. The left two figures are for $n=2$, and right two figures are for $n=6$.}		
		\label{tre2phi*n2}
	\end{center}
\end{figure}
The basic underlying assumption of the above equation is the conservation of reheating entropy during 
the evolution from the radiation dominated phase to the current phase. $g_{re}$ is the number of
relativistic degrees of freedom after the end of reheating phase. 
We also use the following relation between the two temperatures, $T_{\nu 0} = (4/11)^{1/3} T_0$. 
For further calculation, we define a quantity, $\gamma = N^2_{re}/N^1_{re}$. If we identify
the scale of cosmological importance $k$ as the pivot scale of PLANCK, so that $k/a_0 = 0.05 Mpc^{-1}$, and
the corresponding estimated scalar spectral index $n_s = 0.9682 \pm 0.0062$, we arrive at the following 
equation for the efolding number during reheating period, and the reheating temperature,
\begin{eqnarray} \label{nretre}
&& N_{re} = \frac{4(1+\gamma)}{(1-3w_{re1})+\gamma(1-3 w_{re2})}\left[61.6 - \ln\left(\frac{V_{end}^\frac{1}{4}}{H_k}\right) -N_k\right]\\
&& T_{re} = \left[\left(\frac{43}{11g_{re}}\right)^\frac{1}{3} \frac{a_0 T_0}{k} H_ke^{-N_k}\right]^{\frac{3[(1+w_{re1})+\gamma(1+w_{re2})]}
{(3w_{re1}-1)+\gamma(3 w_{re2}-1)}}
 \left[\frac{3^2.5V_{end}}{\pi^2g_{re}}\right]^{\frac{1+\gamma}{(1-3w_{re1})+\gamma(1-3 w_{re2})}}  .
 \label{nretre}
\end{eqnarray}

In the above derivation, we have used $g_{re} = 100$. 
Before discussing any further, let us provide the general descriptions of the figures we have drawn in this section.
As has been mentioned before, we have considered specific values of equation of
state parameter $(w^1_{re},w^2_{re}) =((n-2)/(n+2), 1/3)$ in compatible with our model
discussed in the previous section. It is important to 
mention regarding the special point in the aforementioned state space $(1/3,1/3)$ which is realized for $n=4$. 
Analytically one can check that at this special point both $(T_{re}, N_{re})$ become indeterminate seen in Eq.(\ref{nretre}). 
This fact corresponds to all the vertical solid red lines in $(n_s~vs~T_{re})$ and $(n_s~vs~N_{re})$ plots in Figure \ref{tre1}. 
We have considered $\gamma =1$ as our arbitrary choice. Each curve corresponds
to different values of $n$. On the same plot 
of $(n_s~vs~N_{re})$, we also plotted $(n_s~vs~N_k)$ corresponding to the dotted curves for different models.
At this stage, we would like to remind the reader again that for a wide range of $\phi_*$, all the models predict very small
value of tensor to scalar ratio $r$. Therefore, we will be discussing all the constraints without explicitly mentioning $r$. 
Given the overall description of all the plots, we now set to discuss the
prediction and constraints for two different models.  
In the table-(\ref{tab3}) we provide the important numbers for reheating temperature and the 
e-folding number. We provided only the limiting values of $T_{re}$ which are still allowed from the 
cosmological observation. 
 
\begin{table}[t]
	\begin{tabular}{|p{0.6cm}|p{1.2cm}|p{1.5cm}|p{0.7cm} |p{0.7cm}|}
\hline
		n & $n_s$ & $T_{re}$(GeV) & $N_{re}$ & $N_{k}$ \\
		\hline
		\hline
		2  & \begin{tabular}{c}0.9723 \\0.9702 \end{tabular} &  \begin{tabular}{c} $1\times10^{15}$\\ $1\times10^{3} $ \end{tabular}&
		\begin{tabular}{c} 0.4\\32\end{tabular} & \begin{tabular}{c}54\\50\end{tabular} \\
		\hline
		6  &\begin{tabular}{c} 0.9670\\0.9679 \end{tabular}&\begin{tabular}{c} $1\times10^{14}$\\ $2\times10^{3} $ 
		\end{tabular}& \begin{tabular}{c}00\\23\end{tabular} &\begin{tabular}{c} 53\\55 
	\end{tabular} \\
	\hline
	8  & \begin{tabular}{c}0.9659\\0.9673 \end{tabular}&\begin{tabular}{c} $7\times10^{13}$\\ $1\times10^{3}$\end{tabular} &\begin{tabular}{c} 
		0.6\\23 \end{tabular}&\begin{tabular}{c} 53\\55 \end{tabular}\\
	\hline
	30  & \begin{tabular}{c}0.9625\\0.9653 \end{tabular}& \begin{tabular}{c} $6\times10^{13}$\\ $1\times10^{3} $ \end{tabular}&\begin{tabular}{c} 
		00\\21 \end{tabular}&\begin{tabular}{c} 52\\56 \end{tabular} \\
	\hline
\end{tabular}
\caption{\scriptsize Some sample values of $(n_s, T_{re}, N_{re},N_k)$ are give for two different models
	for $n=(2,6,8,30)$. As we have mentioned, for $n=4$, $(T_{re}, N_{re})$ become indeterministic.
	All these prediction are for $\phi_* = 0.01 {\rm M_p}$.}
\label{tab3}
\end{table}

From the figure we see that for a very small change in $n_s$, the variation of
reheating temperature is very high. Therefore, reheating temperature provides tight constraints on the possible values of e-folding number $N_{re}$. Except for $n=4$, if we restrict the value of $T_{re} \gtrsim 10^3$ GeV, the e-folding number turned out to be $N_{re} \lesssim 35$ during reheating. As an example, for $n=2$, we find spectral index lies within $0.9723 \lesssim n_s \lesssim 0.9702$. Within this range of spectral index, the reheating temperature has to be within $1\times10^{15} \gtrsim T_{re} \gtrsim  1\times10^{3}$
in unit of GeV. This restriction in turn fixed the possible value of e-folding number within a very narrow range $50< N_k < 54$ for $n=2$. For other value of $n$, the ranges are provided in the table-\ref{tab3}. Through our present  analysis one of the important points we infer that in order to achieve the present CMB scale, it must exits the horizon within a very narrow range of e-folding number. This restriction is originated from the allowed range of reheating temperature. The present reheating analysis essentially connects between those two ranges.
	Interesting relation can be found between the reheating temperature $T_{re}$ and the spectral index $n_s$ by numerical fitting them as, 
	 \bea
	 \ln\left( T_{re}\right)\propto A + B(n_s - 0.962) + C(n_s - 0.962)^2 . 
	 \label{nsTrefor}
	 \eea
Where, the dimensionless constants can be approximately found out to be $A=-5\times10^1$, $B =4\times10^3$ and $C=2\times 10^5$ for $n=2$, $A=2\times10^2$, $B =-1\times10^4$ and $C=-7\times 10^5$ for $n=6$ and $A=2\times10^2$, $B =-9\times10^3$ and $C=-4\times 10^5$ for $n=8$. The proportionality constant $Q_p$ is $\phi_*$ dependent constant. Another interesting observation is that with increasing inflaton equation of state, maximum e-folding number during reheating,  $N_{re}$, decreases for a fixed value of reheating temperature. As given in the table, taking  $n=(2,6,8,30)$, the associated maximum $N_{re}$ assumes $(32,23,23,21)$ respectively. This essentially suggests that with increasing inflationary equation of state $w$ or in other words as increase $n$, the faster will be the thermalization process, and consequently earlier will be the radiation dominated phase. To complete the discussion, in the Fig.\ref{tre2phi*n2}, we have also plotted the dependence of reheating parameters for different values of $\phi_*$. We have plotted for $n=2,6$. For all the other models qualitative behaviors of those plots will be same, except $n=4$. Qualitative behavior of $(N_{re},T_{re},N_k)$ remain same for different $\phi_*$. \textcolor{red}{} 
 
\section{\label{conclusion}Summary and Conclusion}
Before we conclude, let us summarize the main results of our study. We studied in detail a specific class of supergravity inspired inflationary models. Most importantly to best fit the experimental data all the important scales $(\phi_*, m)$ which control the inflationary as well as reheating dynamics turned out to be sub-Planckian. Therefore, our model prediction can be trusted from the effective field theory point of view. As a result we also have sub-Planckian field excursion during inflation, which is thought to be important in order to avoid any unwanted quantum gravity effect. However, depending upon the choice of scale we realized both large field as well as small field inflation.
For both the cases, the prediction
of tensor to scalar ratio $(r)$ turned out to be significantly small for a wide range of parameter space. Another important point we would like to point out that our model belong to different class as compared to recently proposed $\alpha$-attractor models.

At this point let us point out that a detailed statistical analysis on inflationary models is done in \cite{Martin:2013tda,martin2} to select the \textit{best fit} models  out of a large number of models. Similar analysis is beyond the scope of the present work. However, an important conclusion of their work is that there exist a common feature among all the best fit models. They found that all these best fit model potentials have a large plateau region, i.e., $V'(\phi)\to 0$ as $\phi\to\infty$. Thus it can be inferred that our power-law plateau model as well as the full supergravity model will also be among the best fit models of inflation.

Near the minimum, the inflaton potentials behaves like a power law $V \propto \phi^n$. Therefore, after the end of inflation coherently oscillating inflaton can be well approximated as an effective fluid with the equation of state parameter, $w=(n-2)/(n+2)$ at least at the initial stage of reheating. With the usual power law potential, one finds it very difficult to fit the cosmological observation for even $n \geq 2$. As a result, detailed studies have not been done for general power law potential specifically in the context of reheating. However, as we have advertised throughout our paper, with the non-polynomial generalization we can easily have power law form specifically at the minimum of the potential. This fact leads us to generalize the work of reheating constraint analysis proposed in \cite{kamionkowski,liddle} for general effective inflaton equation of state $w$. Main outcome of this analysis can be seen in the Table-\ref{tab3}.

After studying the PLANCK constraints it would be very much important to study in detail the reheating dynamics. Reheating is the stage during which all the matter fields are produced. Usual strategy would be to consider the effective coupling among the inflaton and various other matter fields we see today. Considering perturbative reheating phenomena, we have discussed important issue of constraining the dark matter parameter space with CMB data in\cite{Maity:2018exj}. The non-perturbative particle production or preheating for this model with full numerical lattice simulation has been studied in\cite{Maity:2018qhi}.
  
\section{Acknowledgement}
We would like to thank our HEP and Gravity group members for their valuable comments and discussions. We thank Abhijit Saha for useful discussions. We thank the anonymous referees for useful comments which helps us to improve the work.

\appendix

\section{\label{appendix:appA}Towards derivation of $V_{\rm min}$ from non-minimal scalar-tensor theory}

  In this section starting from non-minimal scalar-tensor theory, we will try to construct our model potentials which were a priori ad hoc in nature. As is well known, inflationary models based on power law potential $V(\phi) \sim \phi^n$ are simple but have been ruled out in general because of their large prediction of tensor to scalar ratio. Moreover, the models with large plateaus (Starobinsky or $\alpha$-attractors) are found to be most favored form the PLANCK observation. While most of these plateau models can be cast into exponential potential, plateau potentials with power-law form have also been discussed in super gravity\cite{Dimopoulos:2014boa, Dimopoulos:2016zhy} and non-minimal coupling to gravity\cite{Eshaghi:2015rta, Broy:2016rfg}. In this section we will try to construct our model based on this non-minimally coupled scalar-tensor theory. 
We will see, how simple power-law potentials in the Jordan frame can give rise to the plateau potentials of 
desired form in the Einstein frame. However, this transformed models coincide with our minimal models 
only in a limiting regime (weak conformal coupling). At this point
let us point out that equivalence between the Einstein frame and Jordon frame is an important 
question to ask. This issue has been discussed \cite{Futamase:1987ua, Kaiser:1994vs, Hwang:1996np, Deruelle:2010ht, Postma:2014vaa}, from theoretical as well as cosmological point of views. 

Nevertheless, our motivation in this section is to construct our desired form of the potentials which 
we have shown to be in different class of models rather tan $\alpha$ attractor model. 
We start with the following non-minimally coupled scalar-tensor 
theory, 
\be
S_J = \int d^4x \sqrt{-g} \left[ \frac{\Omega(\varphi)}{2} {\rm M_p^2} R - \frac{\omega(\varphi)}{2} g^{\mu \nu } \partial_{\mu}\varphi \partial_{\nu}\varphi - V(\varphi)     \right],
\ee
where, $\Omega(\varphi), \omega(\varphi)$ are arbitrary function of a scalar field $\varphi$. We will chose a specific form of those
function for our later purpose.
To get the action in the Einstein frame, one performs the following conformal transformation as,
\be
\tg_{\mu \nu} = \Omega(\varphi) g_{\mu \nu} ,
\ee
The action in the Einstein frame can be written as\cite{Fujii:2003pa}
\be
S_E = \int d^4x \sqrt{-\tg} \left[\frac{\rm M_p^2}{2} \tR -\frac{1}{2} F^2(\varphi) \tg^{\mu \nu} \partial_{\mu}\varphi \partial_{\nu}\varphi - \tV(\varphi)      \right]
\ee
	Where, we have assumed that $\omega(\varphi) = \Omega(\varphi)$ and $F$ and the new potential can be found to be,
\be
F^2(\varphi) = \frac{3 {\rm M_p^2}}{2} \frac{\Omega'^2(\varphi)}{\Omega^2(\varphi)} + 1 ~~;~~\tV(\varphi) = \frac{V(\varphi)}{\Omega^2(\varphi)}
\label{Funct}
\ee
Now, we choose the following non-minimal coupling function \cite{Kallosh:2013tua, Galante:2014ifa}, for 
$\Omega^2(\varphi)$, 
\bea
   \Omega^2(\varphi) =
   	1+ \xi (\frac{\varphi}{\rm M_p})^n .
   \label{omega}
   \eea
Therefore, applying (\ref{omega}), we find F and $\tV$ as,
   \bea
   F^2(\varphi) = 
   \frac{3 n^2 \xi ^2 \left(\frac{\varphi }{\rm M_p}\right)^{2 (n-1)}}{8 \left[1 + \xi  \left(\frac{\varphi }{\rm M_p}\right)^n\right]^2}+1~~;~~ \tV(\varphi) = 
   	\frac{V(\varphi)}{1 + \xi \left(\frac{\varphi}{\rm M_p}\right)^n}
   \eea
   
   	We use the following field redefinition 
   	\be
   	\frac{d\phi}{d\varphi} = F(\varphi)
   	\label{redef}
   	\ee
   	to transform the non-minimal into the action of a minimally coupled scalar field with canonical kinetic term,
   	\be
   	S_E = \int d^4 x \sqrt{- \tg} \left[\frac{\rm M_p^2}{2}\tR  -\tg^{\mu \nu } \partial_{\mu}\phi \partial_{\nu}\phi - \tV(\phi)     \right]
   	\ee
	At this point we can integrate Eq.(\ref{redef}), to find the new field in terms of the old field, and construct 
	the modified potential as a function of new field. It is clear from the above set of transformations that
	for entire range of parameter $\xi$, it is very difficult to reproduce our model. 
	However, in the regime of weak coupling $\xi << 1$, $F \sim 1$, hence we can approximately write, using Eq.(\ref{redef}); $\varphi \sim \phi_0 \phi$ ($\phi_0$ is some integration constant). Considering Jordan frame potential as power-law: $V(\varphi) \approx \varphi^n$, one
	gets plateau potential as
	\bea
	\tV(\phi) =
			\frac{\lambda ~m^{4-n} \phi^n}{1 +  \left(\frac{\phi}{\phi_*}\right)^n}
	\eea
where, we identify $\phi_*$ as ${\rm M_p}/\xi^{\frac{1}{n}}$. 
Therefore, in the weak coupling regime, $\xi \ll 1$ or $\phi_* > 1$, the non-minimal scalar tensor theory can give rise
to a large class of minimal cosmologies such as ours which do not belong the $\alpha$-attractor model. 

\section{\label{appendix:appB}Background dependent unitarity}

In this section we briefly discuss about the issue of unitarity. It is obvious that our model is non-renormalizable because of non-linear interaction potential. Hence, in order to be predictive, the cut off scale beyond which our model is non-unitarity should be higher than
the inflationary energy scale $V^{1/4}$. For our convenience we calculated some sample numerical values of the inflation scale given in Tab.(\ref{scales}) for different $n$. From the usual effective field theory point of view, the cut off scale $\Lambda_{c}$ of a model can be extracted from the perturbative coefficient of the potential expanded around zero. Considering the model potential and expanding around zero, one gets,  
\bea
V(\phi) = \lambda \frac{m^{4-n} \phi^n}{1+\left(\frac{\phi}{\phi_*}\right)^n} =
\lambda \sum_{r=0}^{\infty} \frac{(-1)^r \phi^{n(1+r)}}{{m^{n-4}\phi_*}^{n r}} = \lambda \sum_{r'} \frac{{\cal O}^{r'}}{\Lambda_c^{r'-4}}.
\eea
Therefore, we can identify $\Lambda_c =m^{(n-4/(n(r+1)-4))}{\phi_*}^{(n r)/(n(r+1)-4)}$, the scale above which the unitarity will be violated. From the aforementioned expression, it is clear that as we increase the operator 
dimension, the cut off scale decreases. 
Therefore, the lowest value of the cut off scale will be exactly at $\phi_*$ for any value of $n$ in large $r$ limit. From table-\ref{scales}, we can clearly see that all the necessary scales of our model are less than the cut off scale. We have also seen that for smaller values of the scale $\phiast$ the field excursion is also sub-Planckian. Therefore, quantum gravity effect will be unimportant which is generically not true for usual power law inflation. The aforementioned conclusion is from the usual effective field theory point of view. However, the background value of the scalar field may effect the unitarity limit. 

In the following discussion we analyze the aforementioned background dependent unitarity limit. In standard model of particle physics,
it is known that unitarity is dependent upon the Higgs vacuum expectation value. Therefore, this should also be true during
inflation. Following the argument for the Higgs inflation in \cite{Bezrukov:2010jz}, one can compute the background dependent cut off scale $\Lambda(\phi_0)$.
The cut of scale can be read off from the coefficient of the operator of dimension higher than 
four. Therefore, we expand the potential $V(\phi)$ in the inflationary background $\phi_0$ taking $ \phi =\phi_0 + \delta \phi$.
Because of the non-trivial background, we will have all possible operators generated by the expansion. 
The dimension five operator turns out to be 
\bea
\delta V^I = &&
{\cal U}_c\left[50 n (-1 + {\phi_0}^n) (1 + {\phi_0}^n)^3 + 24 (1 + {\phi_0}^n)^4 + 
35 n^2 (1 + {\phi_0}^n)^2 (1 - 4 {\phi_0}^n + {\phi_0}^{2 n}) \right.  \\
&& \left.+ n^4 (1 - 26 {\phi_0}^n + 66 {\phi_0}^{2 n} - 26 {\phi_0}^{3 n} + {\phi_0}^{4 n}) + 
10 n^3 (-1 + 10 {\phi_0}^n - 10 {\phi_0}^{3 n} + {\phi_0}^{4 n}) \right]
\delta \phi^5 \nno
\eea
where,
\bea
{\cal U}_c=\frac{\lambda m^{4-n} \phi*^{n-5} n {\phi_0}^{n-5}}{(120 (1 + {\phi_0}^n)^6) }
\eea 
In the inflationary regime, inflaton usually assuems large field ${\phi_0} \gg 1$ measured in unit of $\phi*$. 
In this limit, the approximate expression for the background dependent cut off turns out to be,  
\bea
\Lambda(\phi_0)=
	\frac {120 m^{n-4} \phi_0^{(5+n)}}{ \lambda \phi*^{n-5} n (24 + 50 n + 35 n^2 + 10 n^3 + n^4) }
\label{powspectrumI}
\eea 
In the inflationary background, one can check that above cut off scale is much larger than $\phi_*$ 
for all possible values of $n$. Therefore, the observable predictions are stable against the
quantum correction. However, detail analysis we will defer for our future studies.

  \hspace{0.5cm}

\end{document}